\documentclass[10pt,twocolumn]{article} 
\usepackage{simpleConference}
\usepackage{times}
\usepackage{graphicx}
\usepackage{amssymb}
\usepackage{amsmath}
\usepackage{algorithm}
\usepackage[noend]{algpseudocode}
\usepackage{algorithm}
\usepackage{algpseudocode}
\usepackage{pifont}
\usepackage{amsfonts}
\usepackage{subcaption}
\usepackage{kantlipsum}       
\usepackage{mwe}  
\usepackage{float}                 
\usepackage{graphicx}
\usepackage{caption}
\usepackage{subcaption}
\usepackage{textcomp}
\usepackage[english]{babel}
\usepackage{amssymb} 
\usepackage{multicol}
\usepackage{graphicx}
\usepackage{dirtytalk}
\usepackage{subcaption}
\usepackage{caption}

\begin{document}

\title{Combined Plant and Controller Design Using Batch Bayesian Optimization: A Case Study in Airborne Wind Energy Systems}

\author{Ali Baheri \\ Chris Vermillion 
\\
University of North Carolina at Charlotte
\\
\\
akhayatb@uncc.edu  \\  cvermill@uncc.edu
}

\maketitle
\thispagestyle{empty}

\begin{abstract}
We present a novel data-driven nested optimization framework that addresses the problem of coupling between plant and controller optimization. This optimization strategy is tailored towards instances where a closed-form expression for the system dynamic response is unobtainable and simulations or experiments are necessary. Specifically, Bayesian Optimization, which is a data-driven technique for finding the optimum of an unknown and expensive-to-evaluate objective function, is employed to solve a nested optimization problem. The underlying objective function is modeled by a Gaussian Process (GP); then, Bayesian Optimization utilizes the predictive uncertainty information from the GP to determine the best subsequent control or plant parameters. The proposed framework differs from the majority of co-design literature where there exists a closed-form model of the system dynamics. Furthermore, we utilize the idea of Batch Bayesian Optimization at the plant optimization level to generate a set of plant designs at each iteration of the overall optimization process, recognizing that there will exist economies of scale in running multiple experiments in each iteration of the plant design process. We validate the proposed framework for Altaeros\textquotesingle\ Buoyant Airborne Turbine (BAT). We choose the horizontal stabilizer area, longitudinal center of mass relative to center of buoyancy (plant parameters), and the pitch angle set-point (controller parameter) as our decision variables. Our results demonstrate that these plant and control parameters converge to their respective optimal values within only a few iterations.
\end{abstract}

\section{Introduction}
Airborne Wind Energy (AWE) systems are a new paradigm for wind turbines in which the structural elements of conventional wind turbines are replaced with tethers and a lifting body (a kite, rigid wing, or aerostat) to harvest wind power from significantly increased altitudes (typically up to 600m). At those altitudes, winds are stronger and more consistent than ground-level winds.

\begin{figure}[t]
    \includegraphics[width=.45\textwidth]{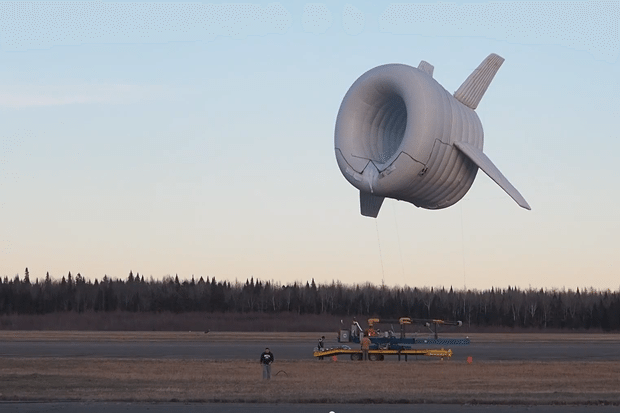}
    \caption{Altaeros Buoyant Airborne Turbine (BAT), Image Credit: \cite{altaeros}}
    \label{fig:BAT}
\end{figure}

The vast energy resource from high-altitude winds has attracted the attention of numerous research and commercial ventures over the past two decades (\cite{altaeros,ampyx,makani,SkySails,vermillion2014model}). To-date, many organizations in the AWE community have focused on optimizing operating altitude (\cite{baheri-acc17,bin2017spatiotemporal}) and crosswind motion to maximize power output, while a limited number of studies have focused on combined plant and control system designs, which have been shown in \cite{nikpoorparizi2016combined} to be coupled. 

Combined plant and controller problems consist of those scenarios in which the optimal controller depends on the physical system design (i.e., the plant) and vice versa. Combined plant and controller optimization (often termed \emph{co-design}) has been employed for a wide variety of systems, including automotive suspension systems (see \cite{fathy2003integrated}, and \cite{allison2014co}), elevator systems \cite{fathy2002nested}, and the AWE application at hand (with initial studies reported in \cite{nikpoorparizi2016combined}, and \cite{joe-ifac2017}). Broadly speaking, combined plant and controller strategies fall into four different categories:
\begin{itemize}

\item \textit{Sequential}: A sequential strategy completes the plant and controller optimization problem in successive order. For instance, authors in \cite{peters2011control} use control proxy functions to augment the objective function of the plant, thereby separating the plant and and controller optimization problem into two sub-problems.

\item \textit{Iterative}: The iterative approach fully optimizes the plant design for a given controller,
then optimizes the controller design for a fixed plant, then repeats the cycle (see \cite{youcef1996modeling}, \cite{reyer1999optimal}).

\item \textit{Nested}: A nested optimization approach contains two loops: an inner loop that completes a full optimization of the controller and an outer loop that completes an iteration of the plant optimization (see  \cite{nikpoorparizi2016combined}, and \cite{fathy2003integrated}, \cite{fathy2002nested}, and \cite{fathy2003nested}).

\item \textit{Simultaneous}: In a simultaneous optimization strategy, both the plant and controller optimization problems are carried out at the same time \cite{athan1996note}, \cite{das1997closer}. An efficient decomposition-based variant of simultaneous optimization is proposed in \cite{allison2010combined}.

\end{itemize}

\tabcolsep=0.05cm

\begin{table*}[]
\centering
\caption{Key existing literature on nested co-design}
\label{}
\small{
\begin{tabular}{clll}
\hline
\textbf{Author}                                   & \multicolumn{1}{c}{\textbf{Basic idea}}                                                                                                                                                                                                           & \multicolumn{1}{c}{\textbf{Optimization technique}}                                                                          & \multicolumn{1}{c}{\textbf{\begin{tabular}[c]{@{}c@{}}Restriction on system dynamic\\ and objective function\end{tabular}}}                                                                      \\ \hline
\textbf{Fathy, et.al (2003) \cite{fathy2001coupling}}                      & \begin{tabular}[c]{@{}l@{}}Outer loop seeks to optimize the overall\\ performance by varying the plant design. \\ The inner loop generates the optimal controller for\\ each plant parameter selected by the outer loop.\end{tabular}             & \begin{tabular}[c]{@{}l@{}}Plant: Interior point method\\ Controller: LQR design\end{tabular}                                     & \begin{tabular}[c]{@{}l@{}}- The structure of system dynamics\\ are known.\\  - System of interest is LTI.\end{tabular}                                                   \\ \hline

\multicolumn{1}{l}{\textbf{Desse, et.al (2017) \cite{joe-ifac2017}}}  & \begin{tabular}[c]{@{}l@{}}The co-design problem is solved using optimal\\ design of experiment for the plant optimization\\ and extremum seeking for the control system\\ optimization.\end{tabular}                                             & \begin{tabular}[c]{@{}l@{}}Plant: G-optimal DoE\\ Controller: extremum seeking\end{tabular}                                  & \begin{tabular}[c]{@{}l@{}}- Extremum seeking may get\\ stuck in a local optima.\end{tabular}                                                                                                        \\ \hline
\multicolumn{1}{l}{\textbf{Baheri, et.al (2017) \cite{baheri-DSCC_CO}}} & \begin{tabular}[c]{@{}l@{}}A machine learning variant of nested co-design.\\ Uses Bayesian Optimization as a global\\ optimization tool for the black-box objective\\ function at both levels.\end{tabular}                                        & \begin{tabular}[c]{@{}l@{}}Plant and Controller:\\ Bayesian Optimization\end{tabular}                                       & \begin{tabular}[c]{@{}l@{}}- No restriction on the system\\ or objective function, but the\\ framework requires simulation \\ and/or experiments.\end{tabular}                                                                                                                                                                   \\ \hline
\label{co-design_literature}
\end{tabular}}
\end{table*}

Among these techniques for solving the co-design problem, the nested co-design approach is unique in its ability to leverage critical differences in control parameters (which can be modified \emph{during} experiments) and plant parameters (which, when experimental work is required, can generally only be modified \emph{between} experiments). This sort of approach can therefore be extremely beneficial in complex systems where experiments will ultimately be required.
However, existing literature on nested co-design, which is summarized in Table \ref{co-design_literature}, makes numerous simplifying assumptions (linearity, full state measurement, etc.). The tools used for the plant adjustment are typically local in nature and often unsuitable for complex systems.
Furthermore, most existing literature on nested co-design uses continuous-time optimal control design techniques (such as LQR) for the controller design (\cite{fathy2001coupling}), without taking advantage of online adaptation capabilities. 

Recently, authors in \cite{joe-ifac2017} proposed a nested co-design approach in which the control parameter is adjusted during a simulation/experiment and plant parameters are optimized \textit{between} simulations/experiments. The benefit here is that when the design optimization process involves lengthy simulations or experiments, the ability to adjust controller parameters during the simulations/experiments can significantly reduce the time and cost of the optimization process. The framework in \cite{joe-ifac2017} used G-optimal design of Experiments (DoE) to select a batch of candidate plant parameters that cover the design space while maximizing a statistical information metric. Furthermore, extremum seeking (ES) control was utilized to adjust the control parameter(s) in real-time over the course of experiments/simulations. The authors recently extended their work in \cite{joe2018Jouranl} to consider online adaptation mechanisms that are based on a global statistical information metric. While resolving some challenges, these approaches suffer from two main drawbacks:
\begin{enumerate}
\item Populating the design space with candidate points in order to merely gain the most information about the design space may lead to the evaluation of candidate designs that have very poor associated performance. This will lead to significant effort expanded in characterizing regions of the design space that are unlikely to yield optimal design parameters.  
\item ES only achieves local, rather than global optimality.
\end{enumerate} 


One of the attractive problems in both the control and machine learning communities is optimizing system designs for real-world applications using \emph{scarce data}. This problem has been studied in the context of sequential decision-making problems aiming to learn the behavior of an objective function (called exploration) while simultaneously trying to maximize or minimize the objective (called exploitation). As an efficient and systematic approach for balancing exploration and exploration in a partially observable environment, Bayesian Optimization has been applied to various real-world problems \cite{baheri2017real, baheri-DSCC_CO, baheri-acc17,baheri2017context, baheri-DSCC_OCT,baheri2018iterative, abdelrahman16bayesian, garnett2010bayesian}. In general, Bayesian Optimization aims to find the global optimum of an unknown, expensive-to-evaluate, and black-box function within only a few evaluations of the objective function at hand.

The work presented here is an extension of the recent DSCC conference paper, \cite{baheri-DSCC_CO}, where we introduced a novel machine learning variant of nested co-design framework. In this framework, Bayesian Optimization is used at both the plant and control design levels of a nested co-design framework, in order to efficiently explore the design space. This framework is applicable to a vast array of problems where there is no analytical expression for the performance metric in terms of decision variables. Furthermore, the proposed framework can easily support experiments in which control parameters are adapted during experiments. In this work, we extend our results from \cite{baheri-DSCC_CO} in the following ways:

\begin{itemize}
\item We employ Batch Bayesian Optimization at the plant optimization level to generate a \emph{set} of plant designs at each iteration of overall optimization process, recognizing that there will exist economies of scale in running multiple experiments at once.
\item We conduct a detailed economic assessment to quantify the cost associated with batch and non-batch Bayesian Optimization algorithms.
\end{itemize}

To validate our approach, we focus on the Altaeros Buoyant Airborne Turbine (BAT) (See Fig. \ref{fig:BAT}). Among parameters critical to system performance, we focus on horizontal stabilizer area and longitudinal center of mass relative to center of buoyancy as plant parameters. We take the trim pitch angle as the controller parameter. The rest of this paper is structured as follows: First, we present the fused plant and control optimization methodology that is used in the nested optimization framework. Next, we summarize the plant and controller design optimization of the BAT. Finally, we provide detailed results, demonstrating the effectiveness of the proposed approach.
\begin{table*}[]
\centering
\caption{Description of key variables involved in the combined plant and controller optimization}
\label{design_params}
\begin{tabular}{cc}
\hline
\textbf{Variable}                                      & \textbf{Description}                                                                                                                              \\ \hline
$\textbf{p}_p$                                         & Plant parameter(s)                                                                                                                    \\
$\textbf{p}_c$                                                     & Control parameter(s)                                                                                                                    \\
$\textbf{p}_{c}^{\ast}(\textbf{p}_{p})$                                            & Optimal control parameter(s) for a candidate plant design                                                                                            \\
$\mathcal{B}_j$                                            & Elements (candidate plant designs) in a batch at iteration $j$                                                                                            \\
\multicolumn{1}{l}{$J(\textbf{p}_{c}^{\ast}(\textbf{p}_{p}),\textbf{p}_p)$} & \begin{tabular}[c]{@{}c@{}}Integral performance value while operating at the optimal \\ control parameter(s) and candidate plant design\end{tabular} \\ \hline
\end{tabular}
\end{table*}

\section{FUSED PLANT AND CONTROLLER METHODOLOGY}

\subsection{PROBLEM FORMULATION}

The ultimate goal of this study is to solve the following optimization problem:

\begin{equation}
\begin{aligned}
& \underset{\textbf{p}_p,\textbf{p}_c}{\text{minimize}}
& & J(\textbf{p}_p,\textbf{p}_c) = \int_{0}^{T_{\mathrm{final}}}g\big(\textbf{x}(t);\textbf{p}_p,\textbf{p}_c\big)dt \\
\end{aligned}
\end{equation}
subject to:
\begin{equation}
\dot{\textbf{x}} = f(\textbf{x},\textbf{u}, \textbf{d};\textbf{p}_p,\textbf{p}_c) 
\end{equation}
\begin{equation}
\textbf{p}_{p} \in P, \ \  \textbf{p}_{c} \in C
\end{equation}
where (1) describes the integral cost functional, (2) represents a general dynamic model that governs the system dynamic, and (3) presents hard constraints on plant and control parameters, which are denoted by $\textbf{p}_p$ and $\textbf{p}_{c}$, respectively. $\textbf{d}$ is the state disturbance vector.
This work will focus on co-design processes that are carried out in controlled environments, where the environmental perturbation (manifested by $\textbf{d}$) is consistent between simulations/experiments but nonetheless important.

\begin{figure}[h]
    \includegraphics[width=.5\textwidth]{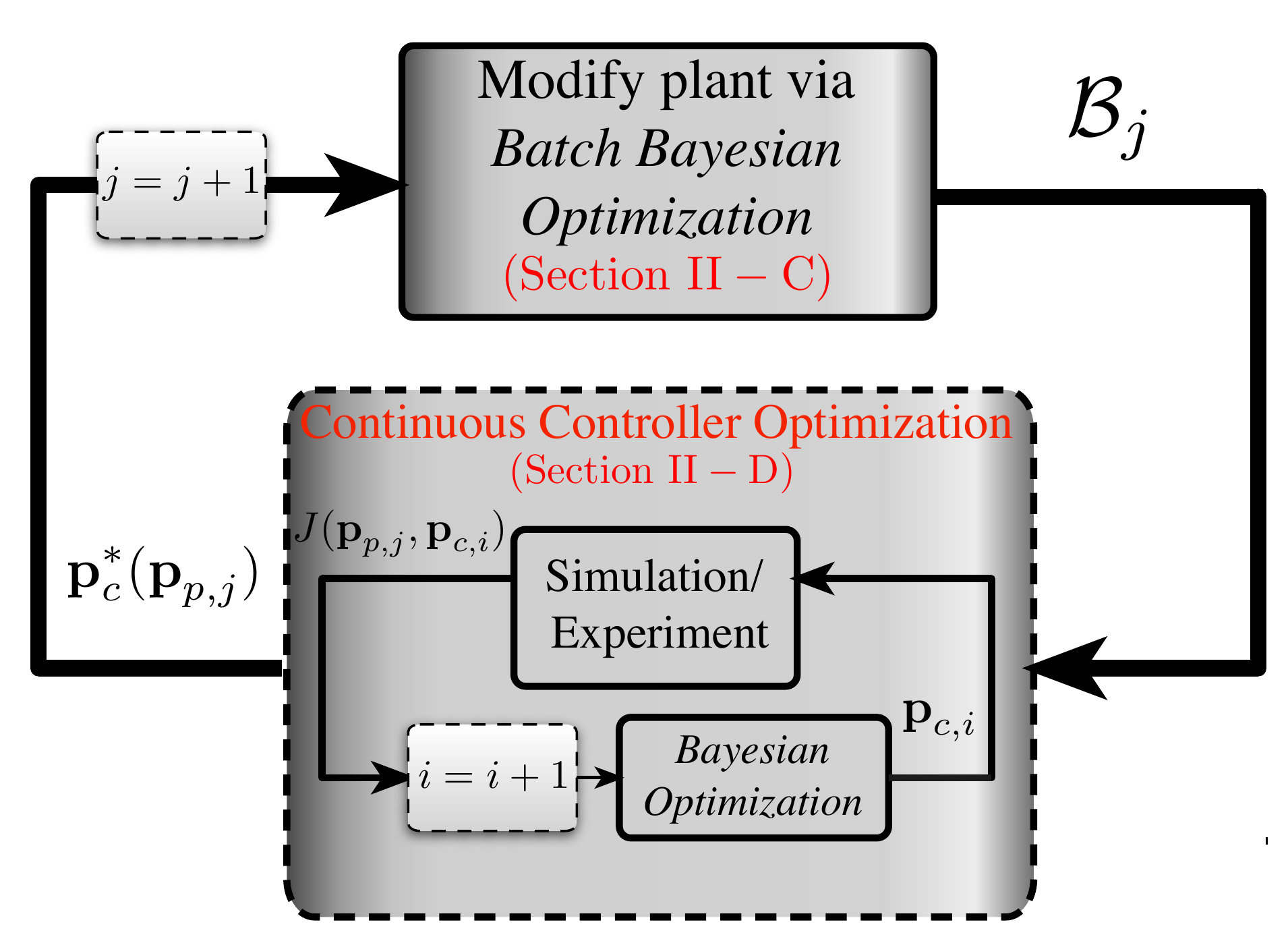}
    \caption{Machine learning variant of nested plant and controller co-design using Batch Bayesian
Optimization}
    \label{fig:framework}
\end{figure}

The nested optimization framework consists of two main loops:
\begin{itemize} 
\item \textbf{Outer loop}: Plant parameters ($\textbf{p}_{p}$) are adjusted in a direction that is chosen by one iteration of our chosen optimization tool (i.e., Bayesian Optimization).
\item \textbf{Inner loop}: For the candidate set of plant parameters generated by the outer loop, the inner loop completes a full optimization of the control parameter vector.
\end{itemize}
Fig. \ref{fig:framework} presents the proposed nested co-design framework. A description of each variable within the general process is provided in Table \ref{design_params}. The process continues until convergence to optimal parameters. Algorithm \ref{alg:CPBO} also summarizes the process proposed in this work.

\begin{algorithm}[t]
\caption{}
\label{alg:CPBO}
\begin{algorithmic}[1]
\Procedure{Fused plant and controller with BO}{} 
\While{plant parameters not converged}
\State Run one iteration of Bayesian Optimization in outer loop
\For {plant candidate parameters}
\While {control parameter not converged}
\State Run full Bayesian Optimization in inner loop
\State Update control parameter
\EndWhile
\EndFor
\EndWhile
\EndProcedure
\end{algorithmic}
\end{algorithm}
\subsection{BAYESIAN OPTIMIZATION}
\label{sec:BO}

The final goal is to optimize the plant and control parameters, (i.e., $\begin{Bmatrix}
\textbf{p}_p,\textbf{p}_c\end{Bmatrix}$), according to some performance index, $J$. Bayesian Optimization is our chosen tool to solve the aforementioned nested optimization problem. Specifically, Bayesian Optimization aims to adjust $\textbf{p}_p$ in the outer loop, while it seeks the optimum of $\textbf{p}_c$ in the inner loop. To illustrate the process of Bayesian Optimization, we take $\textbf{p}$ as the general optimization variable in the rest of this section. Bayesian Optimization traditionally seeks to maximize the value of some performance index. To keep our exposition aligned with this traditional implementation, we define:

\begin{equation}
R   \triangleq -J(\textbf{p})
\end{equation}
as the reward to be maximized.

Bayesian Optimization seeks to globally minimize an unknown, black-box, and expensive-to-evaluate objective function. Due to the computational costs associated with evaluating this function and the possibility of the optimization getting stuck in a local optimum, it is crucial to select the location of each new evaluation deliberately. To address this, Bayesian Optimization is employed to find the \emph{global optimum} of the objective function within \emph{few evaluations} on the real system.
Broadly speaking, there are two main phases in Bayesian Optimization. First, the underlying function is modeled as a Gaussian Process (GP). Second, we choose an acquisition function that guides the optimization by determining the next point to evaluate.
\subsubsection{Learning phase using Gaussian Processes (GPs)}
\label{GP}
In this section, we introduce the basic properties of GPs. GPs are used to model the objective function of the main optimization.

As an attractive choice for non-parametric regression in machine learning, the ultimate goal of GPs is to find an approximation of a nonlinear map, from an input vector to the function value. In general, GP models are able to characterize complex phenomena since they are able to handle the nonlinear effects and interaction between covariates in a systematic fashion.

In general, a GP is fully specified by its mean function, $\mu(\textbf{p})$, and covariance function, $k(\textbf{p},\textbf{p}\textquotesingle)$:
\begin{equation}
R(\textbf{p}) \sim \mathcal{GP}\Big(\mu(\textbf{p}),k(\textbf{p},\textbf{p}\textquotesingle)\Big),
\end{equation}
The GP framework is used to predict the \emph{integral cost function}, $R(\textbf{p})$, at an arbitrary optimization variable(s) (i.e., plant or controller parameters), ${\mathcal{P}}= \begin{Bmatrix}
\textbf{p}_1, \cdots, \textbf{p}_t
\end{Bmatrix}$, based on a set of $t$ past observations, $\mathcal{D}_{1:t} = \begin{Bmatrix}\textbf{p}_{1:t},R_{1:t}(\textbf{p})\end{Bmatrix}$. The function value, represented by $R(\textbf{p}^{\ast })$ in Eq. (\ref{eqn:multivariate_g}), for an unobserved input $\textbf{p}^{\ast }$, and the observed function values, is assumed to follow a multivariate Gaussian distribution \cite{rasmussen2006gaussian}:

\begin{equation}
\begin{bmatrix}
y_{t}\\ R(\textbf{p}^{\ast })\end{bmatrix}  \sim  \mathcal{N}\Big(\textbf{0},\begin{bmatrix}
K_{t} + \sigma_{\epsilon}^{2}{I}_{t}&k_{t}\\ 
 k_{t}^{T}& k(\textbf{p}^{\ast },\textbf{p}^{\ast})
\end{bmatrix}\Big)
\label{eqn:multivariate_g}
\end{equation}
where $y_{t} = \begin{Bmatrix}
R(\textbf{p}_1), \cdots, R(\textbf{p}_t)
\end{Bmatrix}$ is the vector of observed function values. The vector $k_{t}(\textbf{p}) = [k(\textbf{p},\textbf{p}_{1}), \cdots ,k(\textbf{p},\textbf{p}_{t})]$ encodes the relationship between the new input, $\textbf{p}$, and the past data points in $\mathcal{P}$. The covariance matrix has entries $[K_{t}]_{(i,j)} =k(\textbf{p}_{i},\textbf{p}_{j})$ for $i,j \in \begin{Bmatrix}
1, \cdots, t
\end{Bmatrix}$. The identity matrix is represented by ${I}_{t}$ and $\sigma_{\epsilon}$ represents the noise variance \cite{rasmussen2006gaussian}.

The entries in the covariance matrix are parameterized based on a kernel function. A common choice of kernel function is the squared exponential (SE) function, whose evaluation between two inputs, $\textbf{p}_i$ and $\textbf{p}_j$, is expressed as:

\begin{equation}
k(\textbf{p}_i,\textbf{p}_j) = \sigma_{0}^2 \ {\exp} \Big(-\frac{1}{2}(\textbf{p}_i-\textbf{p}_j)^T \Lambda^{-2}(\textbf{p}_i-\textbf{p}_j)\Big),
\end{equation}
where $\gamma = \begin{Bmatrix}\sigma_{0},\Lambda\end{Bmatrix}$ are the hyper-parameters of the kernel function. We select the hyper-parameters as the ones that maximize the marginal log-likelihood of $\mathcal{D}$ \cite{rasmussen2006gaussian}:
 
\begin{equation}
\gamma^{\ast} = \underset{\theta}{\arg \max \log}\  p(y_t\mid \mathcal{P},\gamma),
\label{eqn:hyper-p}
\end{equation}
where
\begin{multline} 
{\log} \ p(y_t\mid \mathcal{P},\gamma) =\\ \Big(-\frac{1}{2}y_{t}^{T}K^{-1}y_{t}-\frac{1}{2}$log$\mid K\mid-\frac{t}{2}$log$2\pi\Big).
\end{multline} 
Once the hyper-parameters are identified, the predictive mean and variance at $\textbf{p}^{\ast }$, conditioned on these past observations, are expressed as:
\begin{equation}
\mu_{t}(\textbf{p}^{\ast}\mid \mathcal{D} ) = k_{t}(\textbf{p})\Big(K_{t}+{I}_{t}\sigma_{\epsilon }^{2}\Big)^{-1}y_{t}^{T},
\label{eqn:gp_mean}
\end{equation}
\begin{equation}
\sigma_{t}^{2}(\textbf{p}^{\ast }\mid \mathcal{D}) = k(\textbf{p},\textbf{p}) - k_{t}(\textbf{p})\Big(K_{t} + {I}_{t}\sigma_{\epsilon }^{2}\Big)^{-1}k_{t}^{T}(\textbf{p}),
\label{eqn:gp_var}
\end{equation}
In summary, the learning phase using GPs involves two main steps: training and prediction. The training step consists of finding proper mean and covariance functions, as well as optimized hyper-parameters, in light of the data (Eq. \ref{eqn:hyper-p}). The prediction phase characterizes the objective function value at an unobserved input in a probabilistic framework (Eqs. \ref{eqn:gp_mean}-\ref{eqn:gp_var}) \cite{rasmussen2006gaussian}. These two equations implicitly serve as surrogates for our unknown function and are used in the next phase to calculate the acquisition function (which is used to determine which point to evaluate next).

\subsubsection{Optimization phase}
As mentioned earlier, we need to choose an acquisition function, which guides the optimization by determining the next point to evaluate. Specifically, the acquisition function uses the predictive mean and variance (Eqs. \ref{eqn:gp_mean}-\ref{eqn:gp_var}) to balance exploration of high-variance regions with exploitation of high-mean regions.

The ultimate goal of Bayesian Optimization is to determine the next optimization variable(s) based on past observations. Among several choices of acquisition functions, we use an acquisition function belonging to the improvement-based family \cite{brochu2010tutorial}. More precisely, the next operating point is selected as the one that maximizes the expected improvement:

\begin{equation}
\textbf{p}_{t+1} = \arg \underset{\textbf{p}} {\max}\ \mathbb{E}\mathbb{I} (\textbf{p}_{t+1}\mid \mathcal{D}_{1:t})
\end{equation}
\noindent where:
\begin{equation}
\mathbb{E}\mathbb{I} (\textbf{p}_{t+1}\mid \mathcal{D}_{1:t}) = \mathbb{E}\Big( \max \begin{Bmatrix}0,R_{t+1}(\textbf{p})-R(\textbf{p})^{\emph{max}} \end{Bmatrix}\mid \mathcal{D}_{1:t}\Big).
\end{equation}
Here, $\max \begin{Bmatrix}0,R_{t+1}(\textbf{p})-R(\textbf{p})^{\emph{max}})\end{Bmatrix}$ represents the \emph{improvement} toward the best value of the objective function so far, $R(\textbf{p})^{\emph{max}}$. The improvement is positive when the prediction is higher than the best value of the objective function so far. Otherwise, it is set to zero. The inability of the acquisition function to assume negative values reflects the fact that if the design \emph{worsens} from one iteration to the next, then it is possible to simply revert to the previous best design. Fortunately, there exists a closed form expression for expected improvement, given by \cite{mockus_ei}: 

\begin{multline} 
\mathbb{E}\mathbb{I} (\textbf{p}_{t+1}\mid \mathcal{D}_{1:t}) = \\ \begin{cases}
\Big(\mu_{t}(\textbf{p})-R(\textbf{p})^{\emph{max}}\Big)\Phi(Z) + \sigma_{t}(\textbf{p})\phi(Z),& \sigma_{t}(\textbf{p}) > 0 \\ 
0, & \sigma_{t}(\textbf{p})= 0
\end{cases}
\label{eqn:EI}
\end{multline} 
where 
\begin{equation}
Z = \frac{\mu_{t}(\textbf{p})-R(\textbf{p})^{\emph{max}}}{\sigma_{t}(\textbf{p})},
\end{equation}
and $\Phi(.)$ and $\phi(.)$ denote the cumulative and probability density function for the normal distribution, respectively.

We summarize the generic version of Bayesian Optimization in Algorithm \ref{alg:BO}. The algorithm is initialized by two previously-evaluated optimization variables and corresponding costs (line 2). Then, at each step, a GP model is trained (line 4) to compute the predictive mean and variance (line 5). These statistical quantities are used to construct the acquisition function (line 6). Next, the point that maximizes the acquisition function is selected as the next candidate (line 7). At the next optimization instance, this point is added to the historical data (line 8), and the process repeats.
\begin{algorithm}[t]
\caption{Bayesian Optimization (BO)}
\label{alg:BO}
\begin{algorithmic}[1]
\Procedure{Generic Bayesian Optimization}{}
\State $\mathcal{D}\gets \textit{Initialize}$$: \begin{Bmatrix}p_{1:2},R(p_{1:2})\end{Bmatrix}$ 
\For{each iteration}
\State Train a GP model from $\mathcal{D}$
\State Compute mean and variance of GP \big(Eqs. (\ref{eqn:gp_mean}-\ref{eqn:gp_var})\big)
\State Compute acquisition function \big(Eq. (\ref{eqn:EI})\big)
\State Find $p^{\ast}$ that optimizes acquisition function
\State Append $\begin{Bmatrix}p^{\ast},{R}(p)\end{Bmatrix}$ to $\mathcal{D}$
\EndFor
\EndProcedure
\end{algorithmic}
\end{algorithm}

\subsection{Bayesian Optimization-Based Co-Design: Batch Plant Design}

The algorithm introduced in our prior conference publication \cite{baheri-DSCC_CO} utilizes Bayesian Optimization to identify just \emph{one} plant design candidate at each iteration of the outer loop. In this section, we examine the use of Batch Bayesian Optimization at the plant optimization level to generate a \emph{set} of plant designs at each iteration of the overall optimization process, recognizing that there will exist economies of scale in running multiple experiments at once.

To mathematically introduce the idea of Batch Bayesian Optimization, we slightly modify the notation of acquisition function presented so far. Specifically, we take $\alpha(\textbf{p}, \mathcal{I}_{t,k})$ as the acquisition function, where $\mathcal{I}$ represents the available data set, $\mathcal{D}$, \emph{plus} the GP structure when $n$ data points are available. Consequently, subscripts $t$ and $k$ represent the $k$th element of $t$th batch, respectfully.

The Local Penalization (LP) algorithm originally presented in \cite{pmlr-v51-gonzalez16a} is a heuristic approach for Batch Bayesian Optimization that works by iteratively penalizing the current peak in the acquisition function to find the next peak. Consequently, each successive element of the batch is chosen to maximize the modified acquisition function, which has been modified based on each previous element of the batch. According to the LP algorithm, every element in $t$th batch is given by:
\begin{equation}
\textbf{p}_{t,k} = \mathrm{arg \  max}_{\textbf{p} \in \textbf{P}}\begin{Bmatrix}
g\big(\alpha(\textbf{p}, \mathcal{I}_{t,0})\big)\prod_{j=1}^{k-1}\varphi(\textbf{p};\textbf{p}_{t,j})
\end{Bmatrix}
\end{equation}
where $g$ is a differentiable transformation of $\alpha$ that keeps the acquisition function positive. $\varphi(\textbf{p};\textbf{p}_{t,j})$ is the core component of LP algorithm and called the \emph{local penalizer} centered at $\textbf{p}_{j}$ (in the $t$th batch). If selected properly, the LP estimates the distance from $\textbf{p}_{j}$ to the true optimum of the cost function. If we believe that the distance from $\textbf{p}_j$ is far from the true optimum, then a large penalizer can discard a large portion of the feasible domain that should not be considered in selecting one of the batch elements. On the other hand, if we believe that $\textbf{p}_{j}$ is close to the true optimum, a small penalizer keeps collecting elements in a close neighborhood of true optimum. However, the main challenge lies in selecting an appropriate local penalizer. 

\subsubsection{Selecting a local penalizer}

As mentioned earlier, the local penalizer characterizes a belief about the distance from the batch locations to true optimum. Let $R_{M} = \mathrm{max}_{\textbf{p} \in \textbf{P}} R(\textbf{p})$. Consider the ball:
\begin{equation}
B_{r_j} (\textbf{p}_j)= \begin{Bmatrix}
\textbf{p} \in \textbf{P}: \left \| \textbf{p}_j - \textbf{p} \right \|\leq r_j
\end{Bmatrix}
\end{equation}
where
\begin{equation}
r_j = \frac{R_{M} - R(\textbf{p}_j)}{L}
\end{equation}
and $L$ is a valid Lipschitz constant of $R$.

If $R(\textbf{p}) \sim \mathcal{GP} \big(\mu(\textbf{p}), k(\textbf{p},\textbf{p}\textquotesingle)\big)$, then we can choose a local penalizer, $\varphi(\textbf{p};\textbf{p}_j)$, as the probability that $\textbf{p}$, any point in $\textbf{P}$ that is a potential candidate for batch elements, does not belong to $B_{r_j} (\textbf{p}_j)$:
\begin{equation}
\varphi(\textbf{p};\textbf{p}_j) = 1- p\Big(\textbf{p} \in B_{r_j}(\textbf{p}_j)\Big)
\end{equation}
However, we need an analytical expression for $\varphi(\textbf{p};\textbf{p}_j)$ to compute each batch element. Proposition 1 provides an analytical form for the local penalizer \cite{pmlr-v51-gonzalez16a}: 
\newline
\textbf{Proposition 1.} \emph{If $R(\textbf{p}) \sim \mathcal{GP} \big(\mu(\textbf{p}), k(\textbf{p},\textbf{p}\textquotesingle)\big)$, then $\varphi(\textbf{p};\textbf{p}_j)$ as defined in (18), is a local penalizer at $\textbf{p}_j$ such that:
\begin{equation}
\varphi(\textbf{p},\textbf{p}_j) = \frac{1}{2}\mathrm{erfc}(-z)
\end{equation}
where
\begin{equation}
z = \frac{1}{\sqrt{2\sigma_n^2(\textbf{p})}}\Big(L\left \| \textbf{p} - \textbf{p}_j \right \| - R_M + \mu_n(\textbf{p}_j)\Big).
\end{equation}
and erfc is the complementary error function}.

Proposition 1 implies that if $\mu_n(\textbf{p}_j)$ is close to $R_{M}$, then $\varphi(\textbf{p},\textbf{p}_j)$ will have a more localized and small impact on $\alpha(\textbf{p})$. On the other hand, if $\mu_n(\textbf{p}_j)$ is far from $R_{M}$, then $\varphi(\textbf{p};\textbf{p}_j)$ will have a big impact on $\alpha(\textbf{p})$.

\subsubsection{Selecting the parameters $L$ and ${R}_{M}$}

An appropriate local penalizer relies on valid choices for $R_{M}$ and $L$. However, the value of $R_M$ and $L$ are unknown in general. To approximate $R_{M}$, one can take 
\begin{equation}
\hat{R}_{M}= \mathrm{max}_i\begin{Bmatrix}
R_i
\end{Bmatrix}
\end{equation}
Regarding the parameter $L$ we take 

\begin{equation}
\hat{L} = \mathrm{max}_{\textbf{p} \in \textbf{P}}\parallel  \nabla{\mu}(\textbf{p})\parallel
\end{equation}
as a valid Lipschitz constant \cite{pmlr-v51-gonzalez16a}.

\begin{algorithm}
\caption{}\label{alg:BBO}
\begin{algorithmic}[1]
\Procedure{Batch Bayesian Optimization with LP}{}
\State \textbf{Inputs:} batch size: $n_b$
\For{$t=1$ until convergence}
\State Fit a GP to $\mathcal{D}_t$
\State Build the acquisition $\alpha(\textbf{p},\mathcal{I}_{t,0})$ using the current GP
\State $\tilde{\alpha}_{t,0}(\textbf{p})\gets g\big(\alpha(\textbf{p},\mathcal{I}_{t,0})\big)$
\State $\hat{L}\gets \mathrm{max}_{\textbf{p} \in \textbf{P}}\parallel \nabla{\mu}(\textbf{p})\parallel $
\For{$j=1$ to $n_b$}
\State $\textbf{p}_{t,j}\gets \mathrm{arg  \ max}  \ ${$\tilde{\alpha}_{t,j-1}(\textbf{p})$}
\State $\tilde{\alpha}_{t,j}(\textbf{p})\gets 
\tilde{\alpha}_{t,0}(\textbf{p})\prod_{j=1}^{k-1}\varphi(\textbf{p};\textbf{p}_{t,j})$
\EndFor\label{euclidendwhile}
\State$ \mathcal{B}^{n_b}_{t}\gets \begin{Bmatrix}
\textbf{p}_{t,1},\dots,\textbf{p}_{t,n_b}
\end{Bmatrix}$
\State $R_{t,1},\dots, R_{t,n_b}\gets \mathrm{evaluation\ of} \  R \  \mathrm{at} \  \mathcal{B}^{n_b}_{t} $ 
\State $\mathcal{D}_{t+1} \gets \mathcal{D}_{t} \  \cup \begin{Bmatrix}
\textbf{p}_{t,j}, R_{t,j}
\end{Bmatrix}_{j=1}^{n_b} $
\EndFor
\State Fit GP to $\mathcal{D}_n$
\EndProcedure
\end{algorithmic}
\end{algorithm}
Finally, algorithm \ref{alg:BBO} summarizes the procedure presented for Batch Bayesian Optimization. 
\subsection{Bayesian Optimization-Based Co-Design: Continuous Controller Optimization}

\begin{figure}[t]
    \includegraphics[width=.47\textwidth]{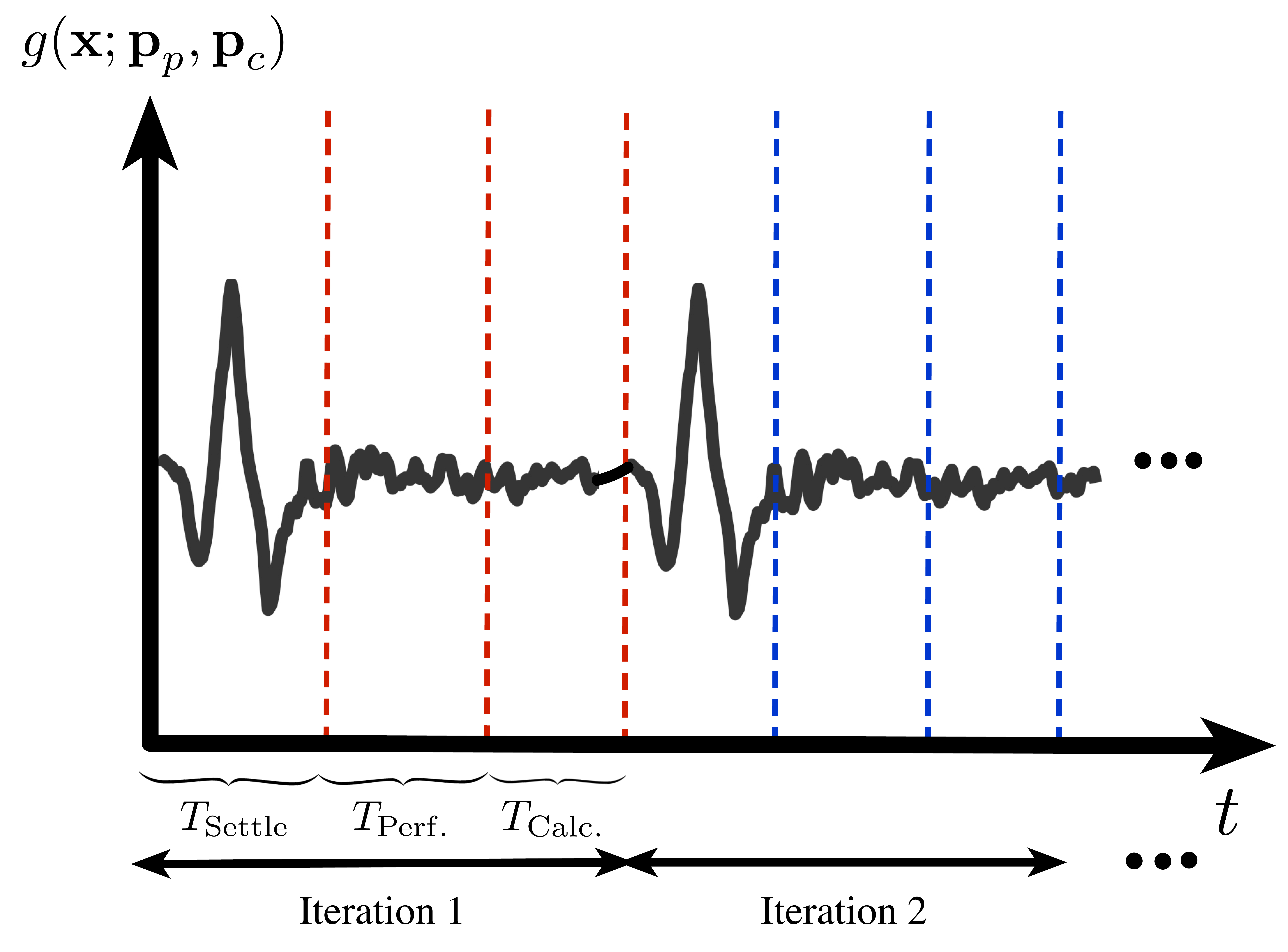}
    \caption{Clarification of iteration vs. time in the proposed framework}
    \label{fig:iteration_concept}
\end{figure}

Bayesian Optimization is an iteration-based optimization algorithm. Thus, in order to use Bayesian Optimization as a tool for adaptation of controller parameters in a continuous-time setting, it is crucial to make a clear connection between the concept of \emph{discrete iterations} (used in Bayesian Optimization), and \emph{continuous time} over the course of a simulation/experiment. Each iteration in the Algorithm \ref{alg:CPBO} corresponds to one window of time within the overall system simulation. We divide each of these windows into three different phases (See Fig. \ref{fig:iteration_concept}). Phase one represents the settling period. To avoid unfairly introducing system transients from one iteration to another in our cost function calculation, we allow the system to settle during the first period, then only compute the cost function value based on the second phase, which we refer to as the performance period. Finally, the calculation of the subsequent decision variable(s) occurs in the third phase, using Bayesian Optimization.

\subsection{Convergence detection}

In any optimization problem, the stopping criteria is determined based on either a fixed iteration budget or certain convergence criteria is met. Detecting convergence, rather than relying on a fixed iteration budget, is critical particularly where experiments come into play, since every experiment requires time and money. In this work we set the following stopping criteria for convergence:

\begin{equation}
|R(\textbf{p}_i) - R(\textbf{p}_{i-j})| < \epsilon  \ \ j = 1,\dots, n
\end{equation}
We set $n$ equal to 2 in this work.
\section{PLANT AND CONTROLLER DESIGN OPTIMIZATION of ALTAEROS BAT}
\subsection{Plant Model of Altaeros BAT}
\label{sec:plant_BAT}

The Altaeros BAT (\cite{altaeros}) features a horizontal axis turbine that is suspended within a buoyant shell. Unlike many AWE concepts, the BAT is designed to remain substantially stationary and passively align with the wind. By accomplishing this, the BAT can achieve secondary objectives, such as telecommunications, which Altaeros has publicly indicated interest in. To that end, it is of great interest to design the combined plant and control system to achieve the steadiest possible flight under atmospheric disturbances, which is the focus of our case study.

\begin{figure}[t]
    \includegraphics[width=.45\textwidth]{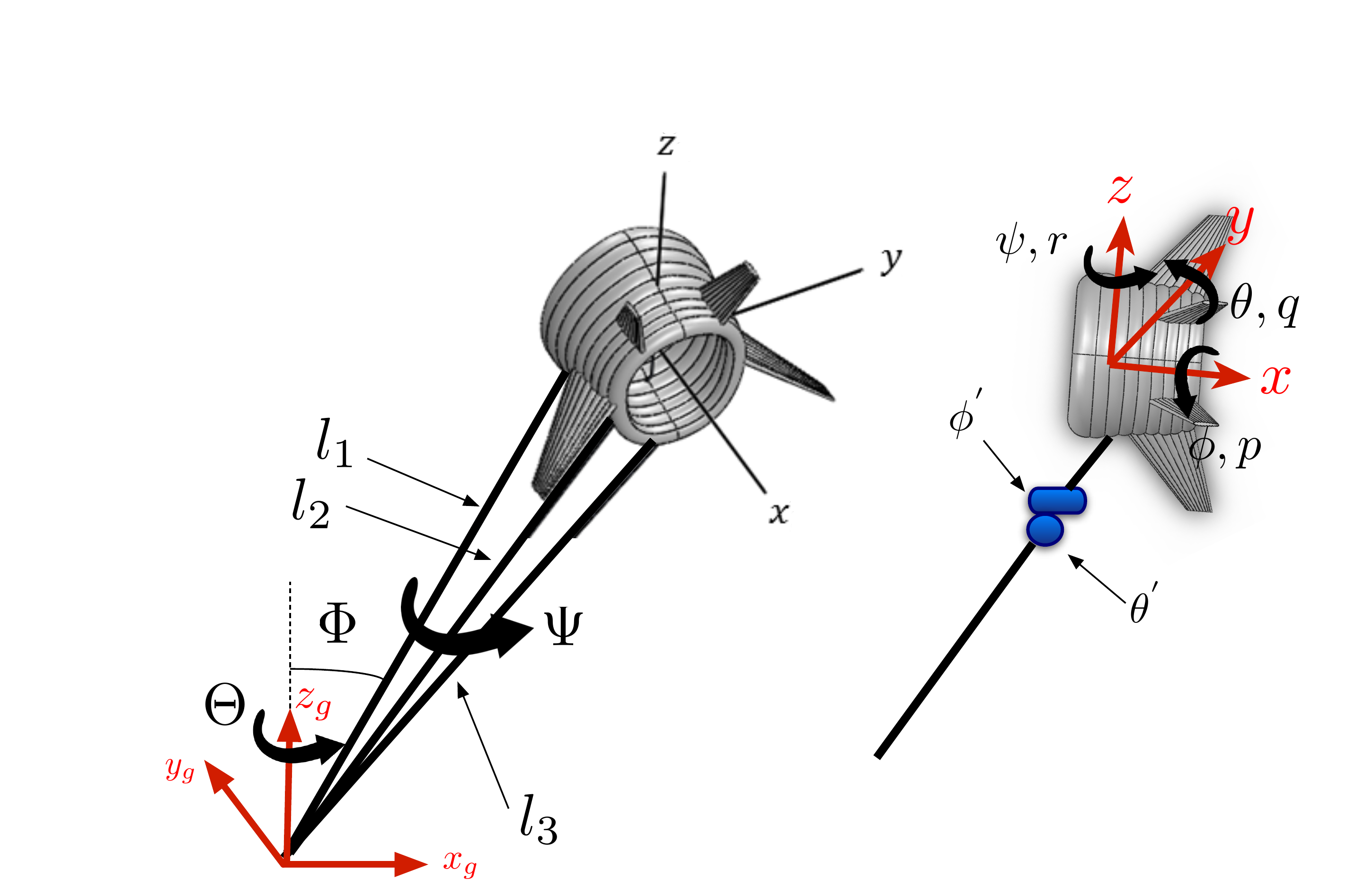}
    \caption{Ground-fixed and body-fixed coordinates plus the key variables used in deriving Euler-Lagrangian dynamics.}
    \label{fig:BAT_Dynamics}
\end{figure}

The dynamic model of Altaeros BAT system was originally introduced in \cite{vermillion2014evaluation}. We briefly introduce some of the important features of this model in this section. Fig. \ref{fig:BAT_Dynamics} shows the variables used in the dynamic model.

The model, which was derived using an Euler-Lagrange framework, describes the position and orientation of the BAT through six generalized coordinates: $\Theta$, $\phi$, $\psi$, $L_t$, ${\theta}^{'}$, and ${\phi}^{'}$. Three of these generalized coordinates represent uncontrolled angles. Specifically, $\Theta$ is the azimuth angle (angle of the tether projection on the horizontal plane), $\Phi$ is the zenith angle (angle of tether with respect to vertical), and $\Psi$ is the twist angle (about the tether axis). Table \ref{full_states} represents the full state variables of AWE system.

The three tethers are modeled as a single tether with length $L_t$ with a bridle joint at the top that splits in three attachment points on the BAT. The bridle joint is modeled through two angles, ${\phi}^{'}$ and ${\theta}^{'}$, which are referred to as induced roll and induced pitch, respectively. The single tether approximation removes algebraic constraint equations, allowing the system to be described by ordinary differential equations (ODEs) rather than differential algebraic equations (DAEs). The center of mass location is modeled as a function of $\Phi$ (zenith angle), $\Theta$ (azimuth angle), and $L_t$ (average tether length). The induced angles are related to the tether lengths ($l_1$, $l_2$, and $l_3$) through the following expressions:
\begin{eqnarray}
\phi' &=& tan^{-1}\Big(\frac{l_{3}-l_{2}}{l^{\mathrm{lat}}_{\mathrm{sep}}}\Big)\\
\theta' &=& tan^{-1}\Big(\frac{l_{1}-0.5(l_{2}+l_{3})}{l^{\mathrm{long}}_{\mathrm{sep}}}\Big)
\end{eqnarray}  
where $l^{\mathrm{long}}_{\mathrm{sep}}$ and $l^{\mathrm{lat}}_{\mathrm{sep}}$ are longitudinal and lateral tether attachment separation distances, respectively. $l_1$, $l_2$ and $l_3$ are the distances between the bridal joint and the three tether attachment points. The control inputs are the tether release speeds, $\bar{u}_i,$ which are given by,
\begin{equation}\label{control input}
\bar{u}_i=\frac{d}{dt}{l}_i
\end{equation}
Ultimately, the governing system equations given below are derived using an Euler-Lagrange formulation and are expressed by:
\begin{equation} 
D(Q)\ddot{Q}+C(Q,\dot{Q})\dot{Q}+g(Q) =\tau\big(Q,\dot{Q},V_{\mathrm{wind}},\psi_{\mathrm{wind}}\big),
\end{equation} 

\begin{eqnarray}
X &=& f(Q,\dot{Q}),\\
\Omega &=& g(Q,\dot{Q}),
\end{eqnarray}

\noindent where:
\begin{eqnarray}\label{physical_params}
Q &=& [\Phi \  \ \Theta \ \ \Psi  \ \ L_t \ \  \theta' \ \  \phi']\\
X &=& [x \ \ y \ \ z \ \ u \ \ v \ \  w]\\
\Omega &=& [\phi \ \ \theta \ \ \psi \ \ p \ \ q \ \ r ]
\end{eqnarray}
\noindent Here, $V_{\mathrm{wind}}$ is the wind speed, and $\psi_{\mathrm{wind}}$ and $\tau$ represent the wind direction and vector of generalized aerodynamic forces and moments, respectively. Aerodynamic forces and moments are functions of $\alpha$ (angle of attack) and $\beta$ (side slip angle), which describe the orientation of the apparent wind vector with respect to the body-fixed coordinates of the BAT. $X$ and $\Omega$ represent the translational and rotational dynamics, respectively. Since we treat the horizontal stabilizer area, $A_{H}$, as a design parameter to be optimized, the aerodynamic coefficients are modeled as explicit functions of the stabilizer areas as follows: 
\begin{center}
\begin{eqnarray}
C_{D,L,S}(\alpha,\beta) &=& C_{D,L,S}^F(\alpha,\beta) + C_{D,L,S}^H(\alpha,\beta)\frac{A^H}{A_{ref}} \nonumber \\
&& +\ C_{D,L,S}^V(\alpha,\beta)\frac{A^V}{A_{ref}} \\
C_{M_x,M_y,M_z}(\alpha,\beta) &=& C_{M_x,M_y,M_z}^F(\alpha,\beta) \nonumber \\
&& +\ C_{M_x,M_y,M_z}^H(\alpha,\beta)\frac{A^Hl^H}{A_{ref}l_{ref}} \nonumber \\
&& +\  C_{M_x,M_y,M_z}^V(\alpha,\beta)\frac{A^Vl^V}{A_{ref}l_{ref}}
\end{eqnarray}
\end{center}
Here, $C_D$, $C_L$, and $C_S$ represent the drag, lift, and side force coefficients, whereas $C_{M_{x}}$, $C_{M_{y}}$, and $C_{M_{z}}$ represent the roll, pitch, and yaw moment coefficients.

\begin{table}[]
\centering
\caption{Full state variables of an AWE system}
\label{my-label}
\begin{tabular}{cc}
\hline
\textbf{State variable}  & \textbf{Notation} \\ \hline
Zenith angle             & $\Phi$            \\
Azimuth angle            & $\Theta$          \\
Twist angle              & $\Psi$            \\
Zenith angle rate        & $\dot{\Phi}$         \\
Azimuth angle rate       & $\dot{\Theta}$       \\
Twist angle rate         & $\dot{\Psi}$         \\
Unstreched tether length & $l$               \\
Induced roll             & $\phi'$           \\
Induced pitch            & $\theta'$         \\ \hline
\end{tabular}
\label{full_states}
\end{table}

\subsection*{Closed-Loop Controller}
\label{sec:flight_controller_BAT}
\begin{figure}[t]
    \includegraphics[width=.47\textwidth]{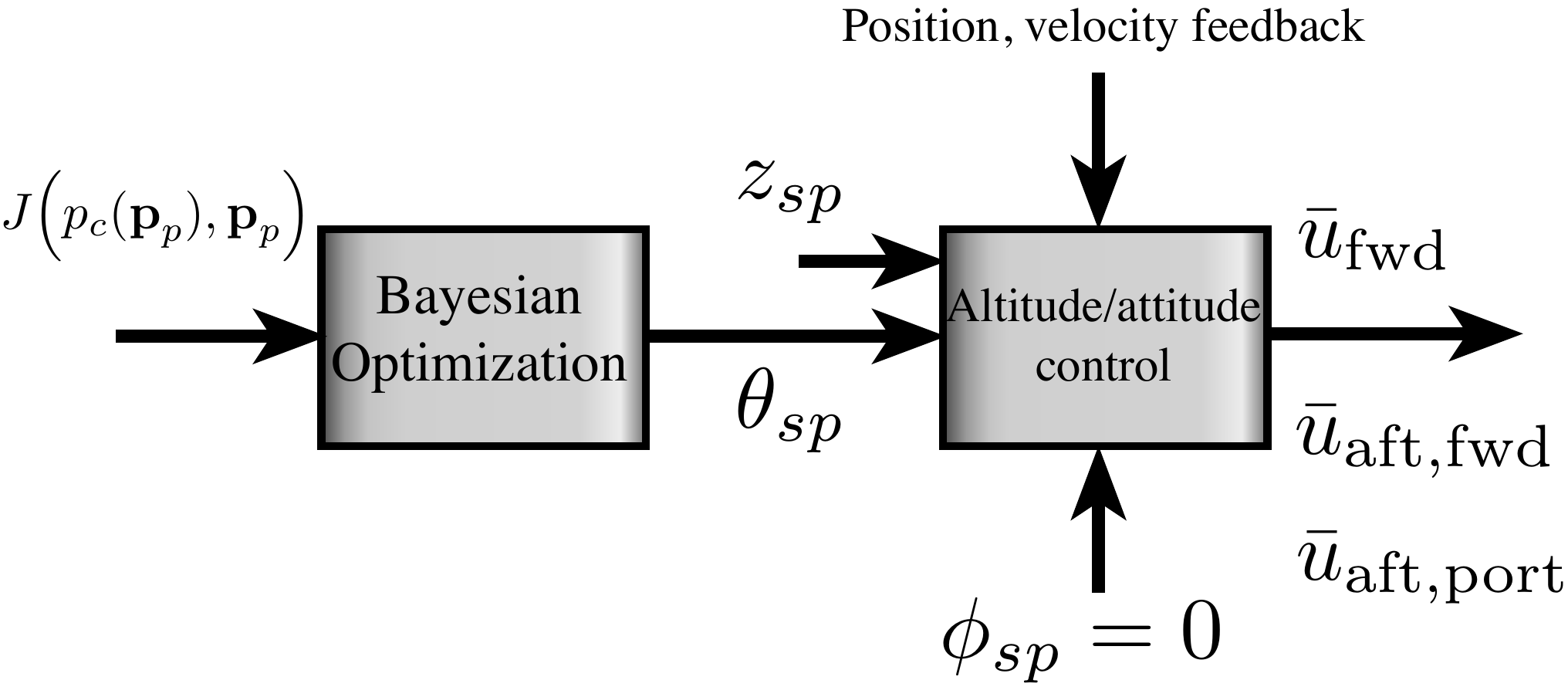}
    \caption{Block diagram of closed loop flight controller for the BAT. $z_{sp}$ denotes a constant altitude set-point. We choose $p_c = \theta_{sp}$ and $\textbf{p}_p =[x_{cm}-x_{cb} \ A_H]^T$ in our case study results.}
    \label{control}
\end{figure}

The controller is designed to track three different set points, namely altitude ($z_{sp}$), pitch ($\theta_{sp}$) and roll ($\phi_{sp}$). Due to the symmetrical configuration, $\phi_{sp}$ is always set to zero. The block diagram of the controller is shown in Fig. \ref{control}. Bayesian Optimization is used to update $\theta_{sp}$, which is computed internally on the flight computer and is not a user-specified value (hence, $\theta_{sp}$ represents a control parameter). Tracking is achieved through the use of three lead filtered PD controllers, which individually control altitude, pitch angle, and roll angle. The outputs of these three lead filters are denoted, respectively, by $\bar{v}_z$, $\bar{v}_{\theta}$, and $\bar{v}_{\phi}$. These outputs represent the average tether release speed (used to regulate altitude), forward/aft tether speed difference (used to regulate pitch), and port/starboard tether speed difference (used to regulate roll), respectively. These controller outputs are related to the tracking errors in altitude ($z_e$), pitch ($\theta_e$) and roll ($\phi_e$) through:
\begin{eqnarray} 
\bar{v}_z(s) &=& \frac{k_{d}^{z}s + k_{p}^{z}}{\tau_{z}s + 1} z_e(s), \\
\bar{v}_{\theta}(s) &=& \frac{k_{d}^{\theta}s + k_{p}^{\theta}}{\tau_{\theta}s + 1} \theta_e(s), \\
\bar{v}_{\phi}(s) &=& \frac{k_{d}^{\phi}s + k_{p}^{\phi}}{\tau_{\phi}s + 1} \phi_e(s).
\end{eqnarray}

\noindent The tether release speeds $\bar{u}_{\mathrm{center}}$, $\bar{u}_{\mathrm{stbd}}$, and $\bar{u}_{\mathrm{port}}$ serve as control inputs to three motors and are related to $\bar{v}_z$, $\bar{v}_{\theta}$, and $\bar{v}_{\phi}$ through:

\begin{equation}
\left[\begin{array}{c} \bar{u}_{\mathrm{center}} \\ \bar{u}_{\mathrm{stbd}} \\ \bar{u}_{\mathrm{port}} \end{array} \right] = \left[\begin{array}{ccc} 1 & -1 & 0 \\ 1 & 1 & 1 \\ 1 & 1 & -1 \end{array} \right] \left[\begin{array}{c} \bar{v}_z \\ \bar{v}_{\theta} \\ \bar{v}_{\phi} \end{array} \right].
\end{equation}

For our case study results, we focus on the following plant parameters to be optimized:

\begin{equation}
\textbf{p}_p = \begin{Bmatrix}
x_{cm}-x_{cb}, &A_{H} 
\end{Bmatrix}
\end{equation}
where $x_{cm}-x_{cb}$ describes the longitudinal center of mass position relative to the center of buoyancy and $A_H$ represents the horizontal stabilizer area.

The controller parameter used in this work is given by:

\begin{equation}
p_c = \theta_{sp}
\end{equation}
where $\theta_{sp}$ presents the trim pitch angle (which is programmed on the flight computer and therefore constitutes a control parameter rather than a user input).
\subsection{PERFORMANCE INDEX} 
The performance index, to be minimized, takes into account two main system properties:
\begin{enumerate}
\item \emph{Ground footprint}: It is of interest to minimize the land usage requirements for multiple systems. For this purpose, the horizontal projected area of land that BAT covers is used as a criterion for quantifying the ground footprint. This area is represented by $A = \pi l^2 \sin^2\Phi$. Thus, as $\Phi$ decreases, the projected area decreases.
\item \emph{Quality of flight}: A low value of heading angle typically corresponds to few oscillations in the system and desirable direct-downwind operation. Furthermore, since we are focused on steady, level flight, we desire to have the BAT as stationary as possible. To characterize the degree to which we accomplish this goal, we penalize heading and roll angle tracking error ($\psi-\psi_{\mathrm{flow}}$ and $\phi-\phi_{\mathrm{sp}}$, respectively) in our performance metric.

\end{enumerate} 
Ultimately, the performance index is denoted by:

\begin{multline} 
J\Big(p_c(\textbf{p}_p),\textbf{p}_p\Big) =\\
 \int_{T_i}^{T_f}\Big(k_1\Phi^2 + k_2(\psi-\psi_{\mathrm{flow}})^2 + k_3(\phi-\phi_{\mathrm{sp}})^2)\Big)dt,
\end{multline} 

\subsection{SIMULATION SETUP}

To excite the system with a wind environment that is consistent across simulations, we implement a frequency approximation of vortex shedding of flow over a cylinder. The model perturbation is based on spectral analysis of flow over a cylinder in \cite{sariouglu2000vortex}. Each component of the velocity is approximated as a sinusoidal perturbation about a mean velocity found in \cite{sariouglu2000vortex}. Each of the velocity components is given by:
\begin{eqnarray}
v_x &=& v_x^{\mathrm{base}}+v_{x0}\sin(\omega_{\mathrm{dist}}t)\\
v_y &=& v_{y0}\sin(\omega_{\mathrm{dist}}t)\\
v_z &=& v_{z0}\sin(\omega_{\mathrm{dist}}t)
\end{eqnarray}  
where $v_\mathrm{base} = 0.606 \frac{m}{s}$, $v_{x0} = 0.0866 \frac{m}{s}$, $v_{y0} = 0.065 \frac{m}{s}$, $v_{z0} = 0.0087 \frac{m}{s}$, and $\omega_{\mathrm{dist}} = 1 HZ$. 

This mechanism for perturbing the system is attractive because it (i) excites key lateral dynamics and (ii) can be implemented in later lab-scale experimental co-design using the team\textquotesingle s water channel setup described in \cite{deodhar2017laboratory} and pictured in Fig. \ref{setup}.

\begin{figure*}[t]
$\begin{array}{rl}
    \includegraphics[width=0.5\textwidth]{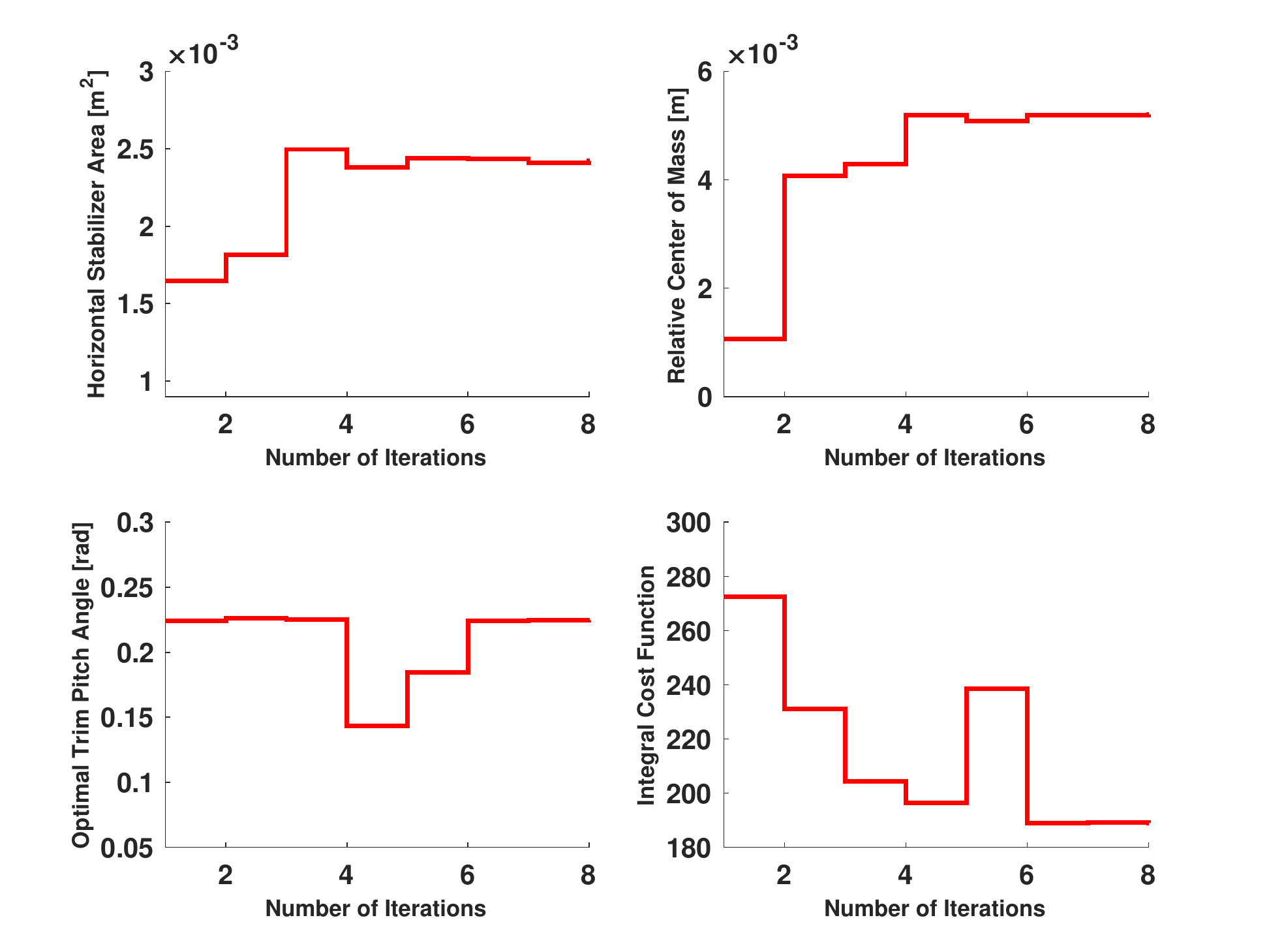}&
    \includegraphics[width=0.5\textwidth]{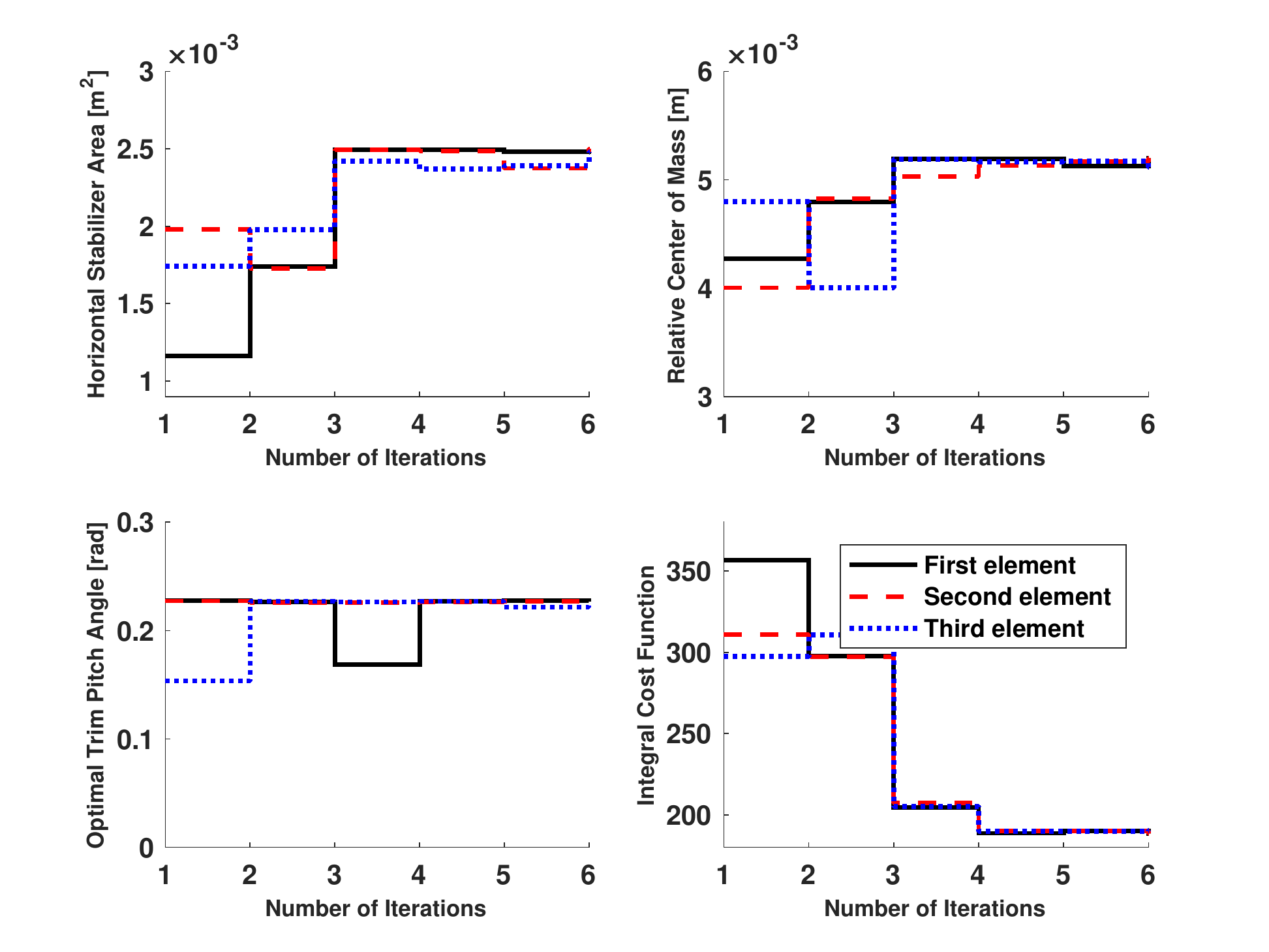}\\
    \multicolumn{2}{c}{\includegraphics[width=0.5\textwidth]{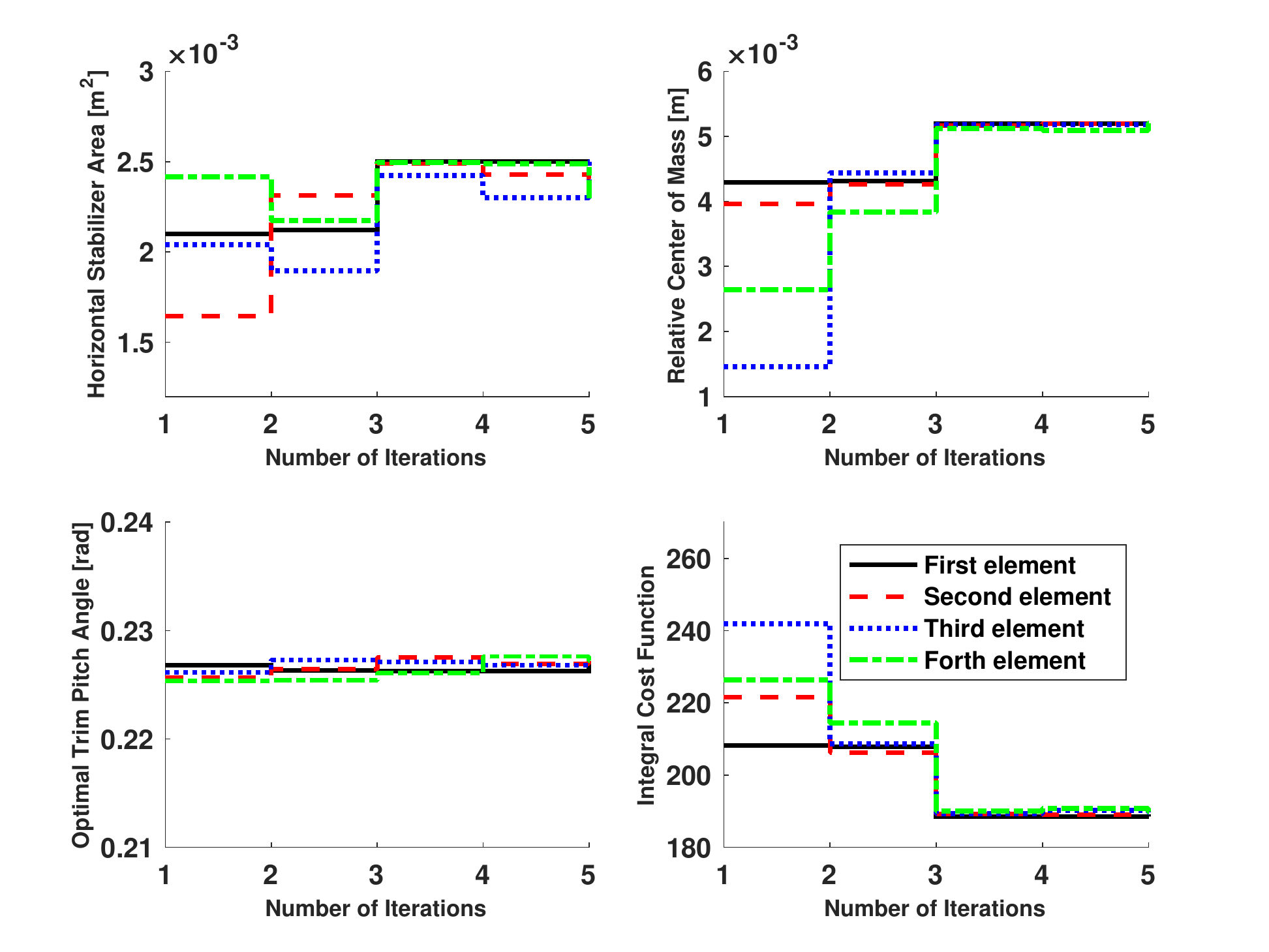}}
\end{array}$
\caption{Convergence of plant parameters, control parameter, and integral cost function for 1, 3, and 4 batch sizes (from upper left to lower right)}
\label{fig:BBO}
\end{figure*}

\begin{figure}[t]
    \includegraphics[width= 0.5\textwidth]{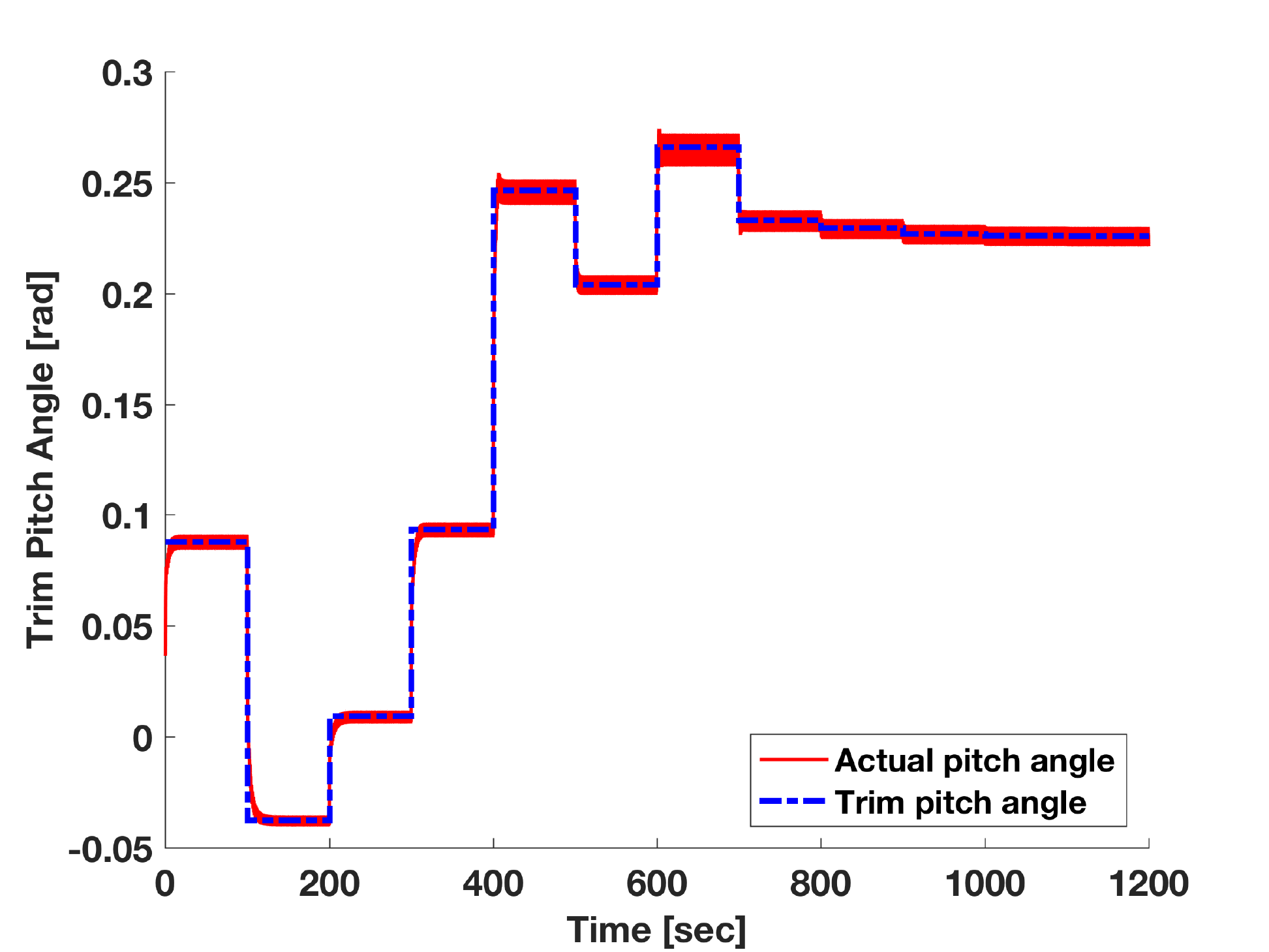}
    \caption{Sample evolution of control parameter (i.e., trim pitch angle) and actual pitch angle in inner loop over the time}
    \label{SamplePitchAngle}
\end{figure}
%
     
\begin{figure*}[t!]
\minipage{0.32\textwidth}
  \includegraphics[width=\linewidth]{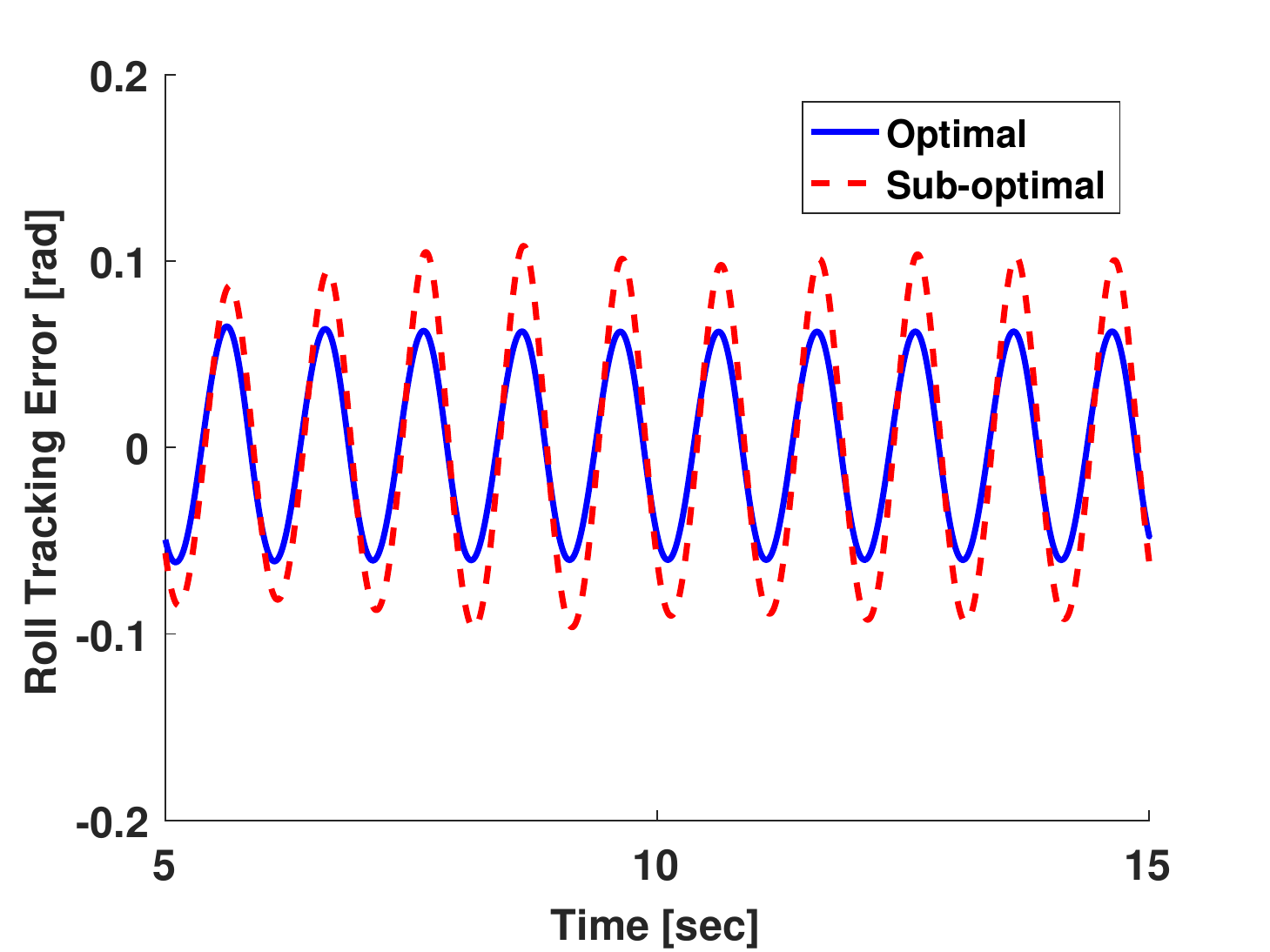}
  \caption{(Zoomed) roll tracking error before and after the optimization}\label{fig:heading}
\endminipage\hfill
\minipage{0.32\textwidth}
  \includegraphics[width=\linewidth]{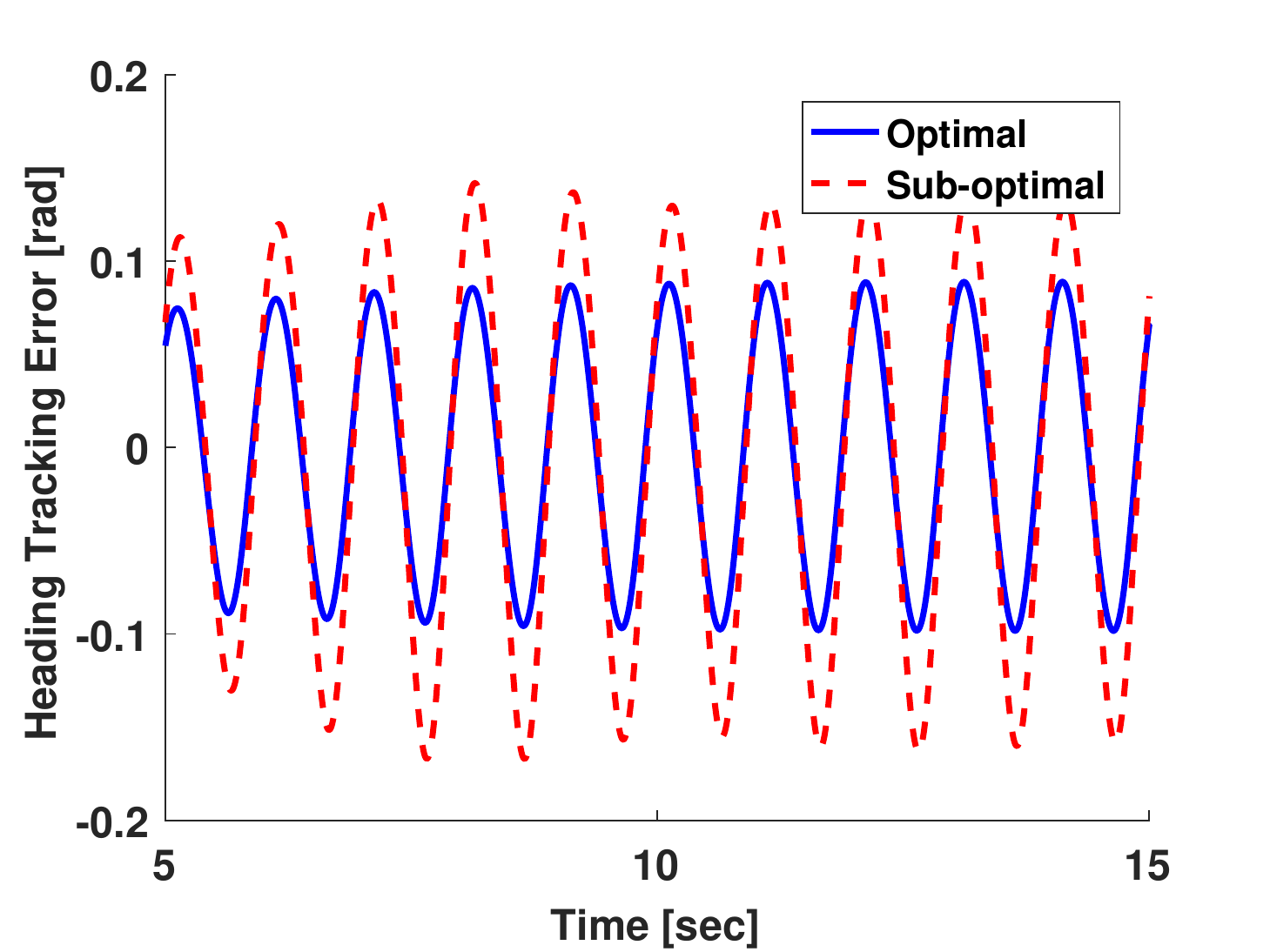}
  \caption{(Zoomed) heading tracking error before and after the optimization}\label{fig:roll}
\endminipage\hfill
\minipage{0.32\textwidth}%
  \includegraphics[width=\linewidth]{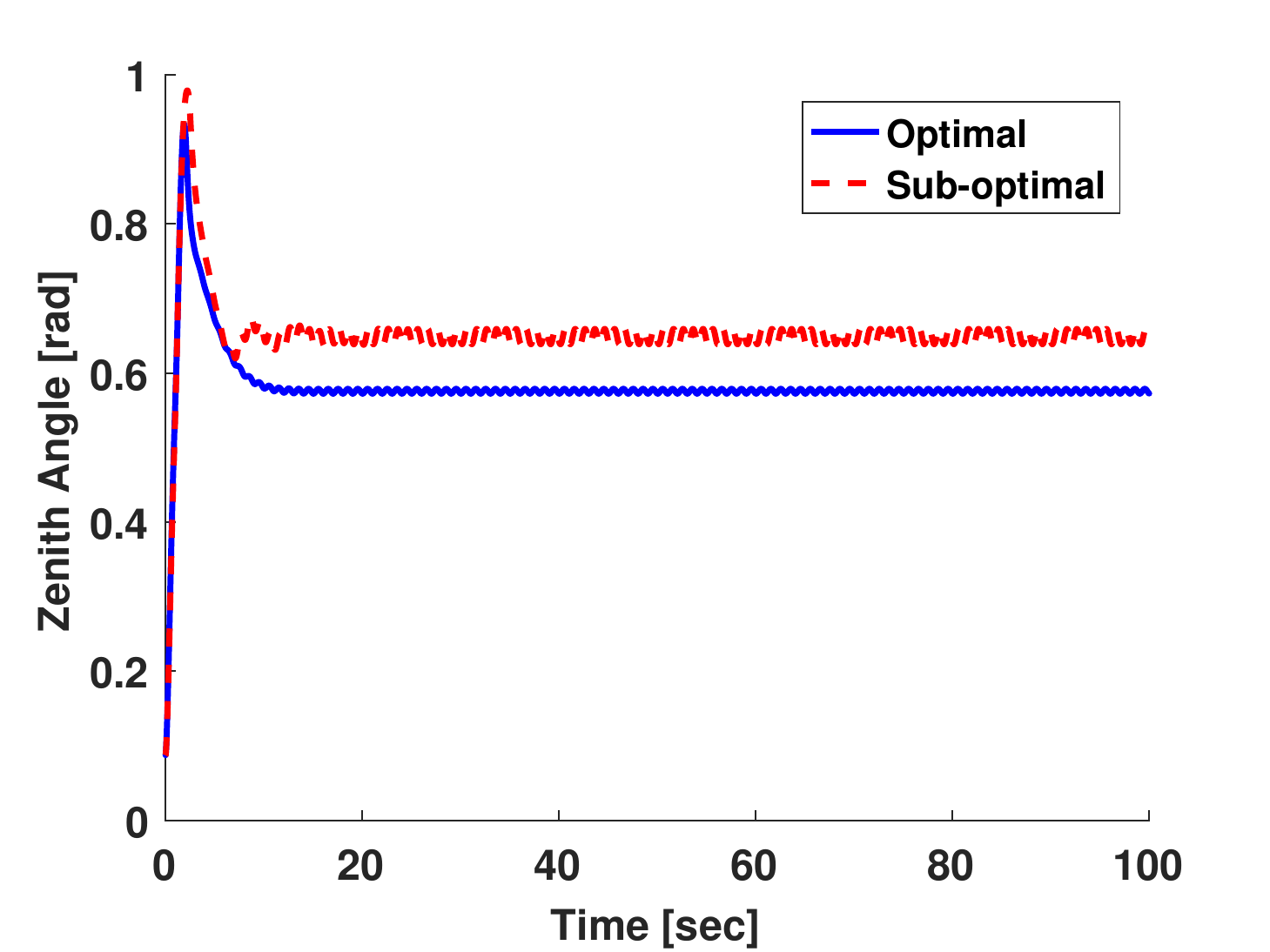}
  \caption{Zenith angle before and after the optimization}\label{fig:zenith}
\endminipage
\end{figure*}

\section{RESULTS}
We evaluated the proposed algorithm on the BAT numerical model. We assessed the effectiveness of Batch Bayesian Optimization algorithm for 3 and 4 elements (plant designs) in each batch. We also compare results from Batch Bayesian Optimization against results for a batch size of 1 (i.e., generic Bayesian Optimization). 

Fig. \ref{fig:BBO} shows the convergence of plant parameters (horizontal stabilizer area and relative center of mass), control parameter (trim pitch angle), and integral of cost function for different batch sizes. Each line structure in this figure represents the evolution of a different batch element over the course of optimization. 

One can immediately see that fewer total iterations are required to converge with larger batch sizes. Furthermore, as can be seen from this figure, the plant parameters converge after only a small number of iterations (each iteration corresponds to one round of system performance evaluation). With the same simulation setup, comparing these results with those reported in {\cite{joe-ifac2017} reveals that Bayesian Optimization leads to a faster convergence than the optimal DoE proposed in that work. 
It should be emphasized that plant design changes are much more expensive compared to adjusting control parameters in instances when simulations are replaced with experiments (which is a long-term goal of the present work). Therefore, reducing the number of required plant reconfigurations in the design optimization process is a very important goal.

Fig. \ref{fig:BBO} also illustrates the optimal control parameter in the inner loop for each plant design generated by the outer loop. Furthermore, Fig. \ref{SamplePitchAngle} represents the sample evolution of control parameter (trim pitch angle) and actual pitch angle in the inner loop over a simulation.

Figs. \ref{fig:heading}, \ref{fig:roll}, and \ref{fig:zenith} illustrate the evolution of the individual instantaneous cost function components (i.e., roll tracking error, heading tracking error, and zenith angle) before and after the iterative optimization process to demonstrate the effectiveness of the proposed approach.

\section{Economies of Scale}
So far, we have explored the idea of Batch Bayesian Optimization at the plant optimization level to generate a set of plant designs at each iteration of the overall optimization process. In this section, we will assess, for the AWE system, whether this will result in economies of scale.

To assess the economies of scale that are introduced through a Batch Bayesian Optimization process, we focus our attention on a lab-scale experimental platform (depicted in Fig. \ref{setup}) for closed-loop flight characterization of AWE systems, which is detailed in \cite{deodhar2017laboratory}. With this system, 3D printed models (depicted in Fig. \ref{scaleBAT}) of AWE system lifting bodies are tethered and flown in the UNC-Charlotte water channel. Micro DC motors are used to regulate tether lengths, high-speed cameras are used for image capture, and a high-performance target computer is used for real-time image processing and control. Because this experimental platform represents the ultimate \say{end game} for the Batch Bayesian Optimization approach, we use it as the basis for the economies of scale analysis presented herein.


In order to run an experimental batch (and non-batch) Bayesian Optimization in water channel, the following steps need to happen:

\begin{figure}[!tbp]
  \centering
  \begin{minipage}[b]{0.3\textwidth}
    \includegraphics[width=\textwidth]{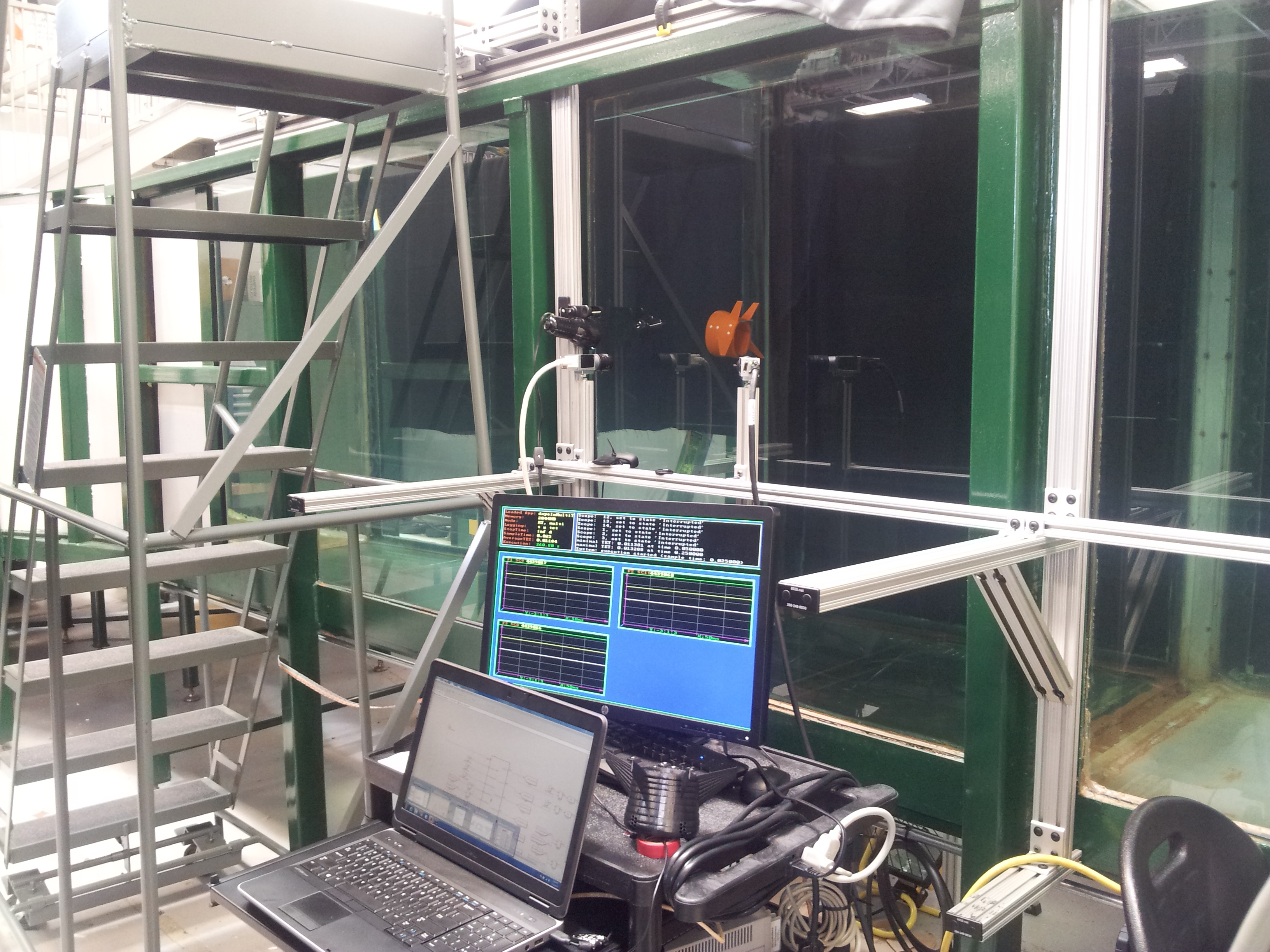}
    \caption{Water channel experimental setup at UNC Charlotte}
    \label{setup}
  \end{minipage}
  \hfill
  \begin{minipage}[b]{0.3\textwidth}
    \includegraphics[width=\textwidth]{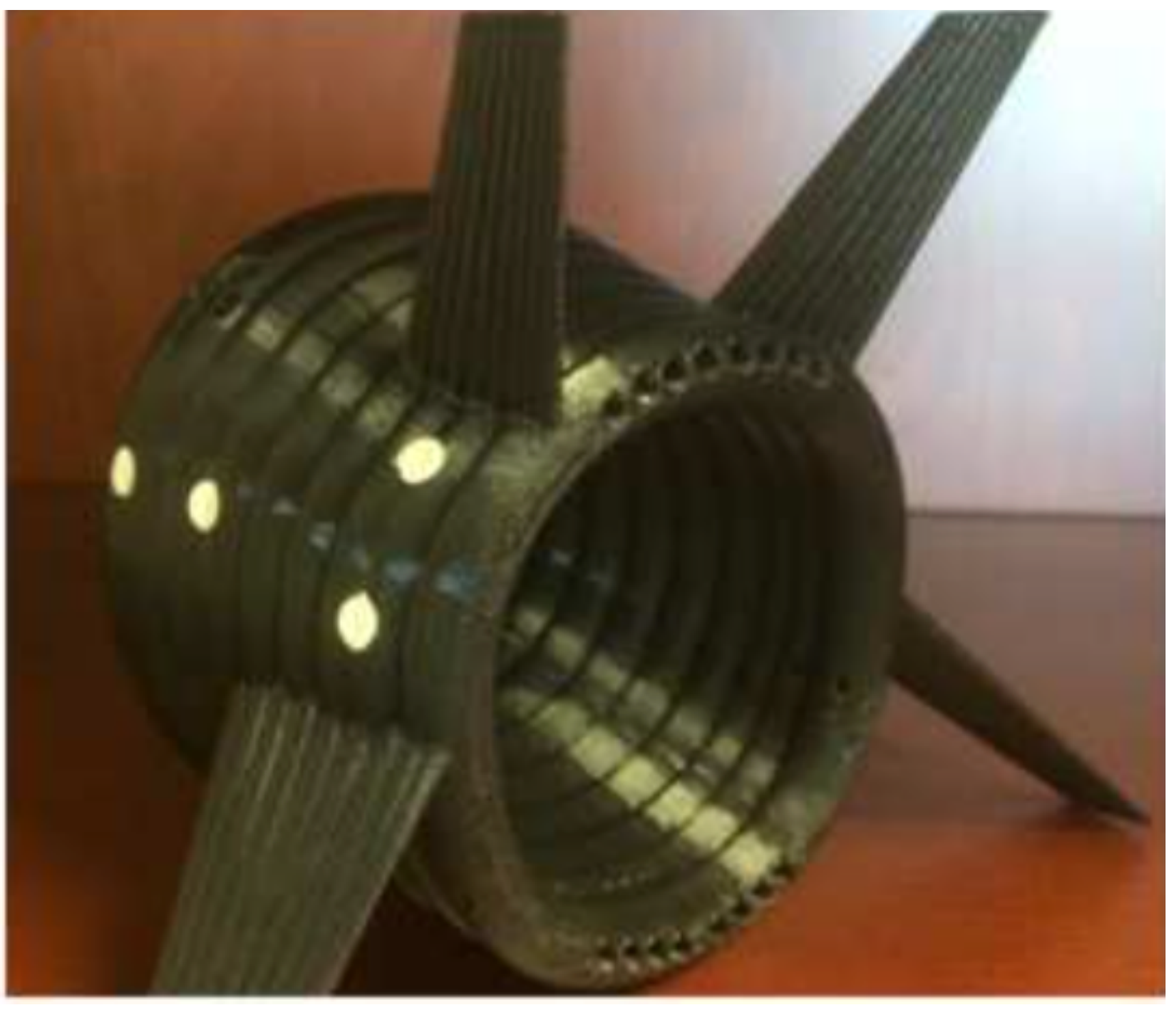}
    \caption{$\frac{1}{100}$-scale BAT model}
    \label{scaleBAT}
  \end{minipage}
\end{figure}

\begin{table*}[]
\centering
\caption{Design parameters for economic assessment}
\label{my-label}
\begin{tabular}{ccc}
\hline
\textbf{Parameter}         & \textbf{Description}                                                      & \textbf{Value}             \\ \hline
$c_{\mathrm{eng}}$    & \multicolumn{1}{l}{cost required to hire employee(s)\
 to conduct experiments} & $\frac{\$30}{\mathrm{hours}}$ \\
$c_{\mathrm{recharge}}$    & cost required for equipment recharge                                      & $\frac{\$240}{\mathrm{day}}$ \\
$c_{\mathrm{Wrecharge}}$    & cost required for water channel recharge                                      & $\frac{\$2400}{\mathrm{day}}$ \\
$c_{\mathrm{lostTime}}$        & opportunity cost of lost time                                                           & $\frac{\$1200}{\mathrm{day}}$ \\
$T_{\mathrm{print}}$       & time required to print the main body                                      & 12 hours                   \\
$T_{\mathrm{3DFins}}$ & time required to print the fins                                           & 4 hours                    \\
$T_{\mathrm{lead}}$        & lead time                                                                 & 5 days                      \\
$T_{\mathrm{setup}}$       & time required to setup experiment                                         & 30 min                     \\
$T_{\mathrm{reconfig}}$    & time required to reconfigure the model                                    & 5 min                      \\
$T_{\mathrm{exp}}$         & time required to conduct one experiment in inner loop                                   & 1200 sec                   \\
$N$                        & number of elements per batch                                              & 1, 3, 4                    \\
$m$                        & number of employee(s) required to print the model and fins                         & 1
\\
${m}'$                        & number of employee(s) required to conduct experiments                         & 2                                     \\ \hline

\end{tabular}
\label{economy}
\end{table*}

\begin{itemize}

\item Prior to running any experiments: 3D print the main body.

The cost associated with this step is given by:

\begin{equation}
C_{3D} = \Big(m\times c_{\mathrm{eng}} + c_{\mathrm{recharge}}\Big) N  T_{\mathrm{print}} 
\end{equation}
where $c_{\mathrm{eng}}$ and $c_{\mathrm{recharge}}$ represent the cost required for engineer(s) to conduct experiments and the facility charge, respectively. $T_{\mathrm{print}}$ denotes the time required to print the main body. $m$ is the number of employees required to print the model (and fins).
%

\item After each batch is chosen, the following needs to happen once per batch:

(i) 3D print $N$ sets of fins. This cost is given by:

\begin{equation}
C_{3D\mathrm{Fins}} = \Big(m\times c_{\mathrm{eng}} + c_{\mathrm{recharge}}\Big) N  T_{3D\mathrm{Fins}} 
\end{equation}
where $T_{3D\mathrm{Fins}}$ represents the time required to print the fins.

(ii) Schedule time in the water channel (this time may include lead time.)

\begin{equation}
C_{\mathrm{lead}} = c_{\mathrm{lostTime}} T_{\mathrm{lead}}
\end{equation}
where $c_{\mathrm{lostTime}}$ and $T_{\mathrm{lead}}$ represent the opportunity cost of lost time and the lead time, respectively.

(iii) Initialize the water channel:

\begin{equation}
\begin{aligned}
C_{\mathrm{Wchannel}} ={} & \Big({m}' \times c_{\mathrm{eng}}+c_{\mathrm{Wrecharge}}\Big) \Big(T_{\mathrm{setup}}      &\\ &+ N T_{\mathrm{exp}}  + N  T_{\mathrm{reconfig}}\Big)
      & \\
\end{aligned}
\end{equation}
where $T_{\mathrm{setup}}$ and $T_{\mathrm{reconfig}}$ denote the time required to set up the experiment in the water channel and the time required to reconfigure the model, respectively. Furthermore, $c_{\mathrm{Wrecharge}}$ and $T_{\mathrm{exp}}$ represent the cost required to run the water and time required to conduct an experiment in the inner loop, respectively. ${m}'$ is the number of employees required to conduct experiments.
\item Between each experiment within a batch, the following needs to happen $N$ times per batch:

(i) Swap out fins; (ii) Adjust center of mass; (iii) Initialize image processing for a given configuration; (iv) Run the experiment.

\end{itemize}

Table \ref{economy} summarizes the design parameters for economic assessment. The total cost can be ultimately expressed as:

\begin{equation}
C_{\mathrm{total}} = N_{\mathrm{convergence}} \times \Big (C_{3D} + C_{3D\mathrm{Fins}} +  C_{\mathrm{lead}} + C_{\mathrm{Wchannel}}\Big)
\end{equation}
where $N_{\mathrm{convergence}}$ denotes the number of iterations required for the convergence based on the number elements per batch. As can be seen from $C_{\mathrm{total}}$, some terms depend on $N$, while some terms do not. This indicates that there exist economies of scale in running Batch Bayesian Optimization. Table \ref{cost} shows the cost associated with different batch sizes according to our economic assessment. One can conclude from this table that as the number of elements per batch increases, the associated cost decreases.

\begin{table}[]
\centering
\caption{Cost in $ \$ $ associated with running experiments with different batch sizes}
\label{cost}
\begin{tabular}{cc}
\hline
\textbf{\begin{tabular}[c]{@{}c@{}}\# of elements \\ per batch\end{tabular}} & \textbf{Cost ($\$$)} \\ \hline
\textbf{1}                                                                   & 54293                   \\
\textbf{3}                                                                   & 49200                   \\
\textbf{4}                                                                   & 44533                   \\ \hline
\end{tabular}
\end{table}
\section{Conclusion}
This work presented a nested co-design framework where Batch Bayesian Optimization replaced local techniques at both the plant and controller optimization levels. Our results using an AWE system demonstrate that both the plant and controller parameters converge to their respective optimal values within only a few iterations. Furthermore, the results confirm that economies of scale exist with Batch Bayesian Optimization approach.

\bibliographystyle{unsrt}
\bibliography{Bo_Summer2016}

\end{document}